\documentclass[12pt,showpacs,preprintnumbers,superscriptaddress,nofootinbib]{revtex4}
\usepackage{amsfonts}
\usepackage{amssymb}
\usepackage{amsmath, graphicx}
\usepackage{dcolumn}
\usepackage{bm}
\usepackage{epstopdf}
\usepackage{color}
\usepackage{soul}
\usepackage{cancel}

\newcommand{\hs}{\hspace*{0.3cm}}
\newcommand{\cm}{\hspace*{1cm}}

\newcommand{\be}{\begin{equation}}
	\newcommand{\ee}{\end{equation}}
\newcommand{\bea}{\begin{eqnarray}}
	\newcommand{\eea}{\end{eqnarray}}
\newcommand{\ben}{\begin{enumerate}}
	\newcommand{\een}{\end{enumerate}}
\newcommand{\bit}{\begin{itemize}}
	\newcommand{\eit}{\end{itemize}}
\newcommand{\bde}{\begin{widetext}}
	\newcommand{\ede}{\end{widetext}}
\newcommand{\nn}{\nonumber}
\newcommand{\crn}{\nonumber \\}

\newcommand{\al}{\alpha}
\newcommand{\la}{\lambda}

\newcommand{\va}{\varphi}
\newcommand{\om}{\omega}

\newcommand{\+}{\dagger}
\newcommand{\fr}{\frac}

\newcommand{\bc}{\begin{center}}
	\newcommand{\ec}{\end{center}}
\newcommand{\Ga}{\Gamma}

\newcommand{\De}{\Delta}

\newcommand{\varep}{\varepsilon}

\newcommand{\La}{\Lambda}
\newcommand{\si}{\sigma}

\newcommand{\Om}{\Omega}
\newcommand{\Long}[1]{{\color{red} #1 }}

\input{tcilatex}

\begin{document}

\title{\boldmath Some phenomenological aspects of the 3-3-1 model with CKS mechanism }
\author{H. N. Long}
\email{hoangngoclong@tdtu.edu.vn}
\affiliation{Theoretical Particle Physics and Cosmology Research Group, Advanced
	Institute of Materials Science, Ton Duc Thang University, Ho Chi Minh City 700000,
	Vietnam}
\affiliation{Faculty of Applied Sciences, Ton Duc Thang University, Ho Chi Minh City 700000,
	Vietnam}
\author{N. V. Hop}
\email{nvhop@ctu.edu.vn}
\affiliation{Department of Physics, Can Tho University, 3/2 Street, Can Tho 900000, Vietnam}
\author{L.T. Hue}
\email{lthue@iop.vast.vn}
\affiliation{Institute for Research and Development, Duy Tan University, Da Nang City 550000,
	Vietnam}
\affiliation{Institute of Physics, Vietnam Academy of Science and Technology, 10 Dao Tan,
	Ba Dinh, Hanoi 100000, Vietnam }
\author{N. H. Thao}
\email{abcthao@gmail.com}
\affiliation{Department of Physics, Hanoi Pedagogical University 2, Phuc Yen, Vinh Phuc 280000,
	Vietnam}
\author{A. E. C\'arcamo Hern\'andez}
\email{antonio.carcamo@usm.cl}
\affiliation{Universidad T\'{e}cnica Federico Santa Mar\'{\i}a and Centro Cient\'{\i}%
	fico-Tecnol\'{o}gico de Valpara\'{\i}so, \\
	Casilla 110-V, Valpara\'{\i}so, Chile}
\date{\today }

\begin{abstract}
We perform a comprehensive analysis of several phenomenological aspects of the renormalizable extension 
of the inert 3-3-1 model 
with sequentially loop-generated SM fermion mass hierarchy. Special attention is paid to the study 
of the constraints arising from the experimental data on the $\rho$ parameter, as well as those ones resulting from the charged lepton flavor 
violating process $\mu\to e\gamma$ and dark matter. We also study the single $Z'$ production via 
Drell-Yan mechanism at the LHC. 
%
 We have found that $Z'$ gauge bosons heavier than about $4$ TeV comply with the experimental constraints
   on the oblique $\rho$ parameter as well as with the collider constraints. In addition, we have found that the 
   constraint on the charged lepton flavor violating decay $\mu\rightarrow e\gamma$ sets the sterile neutrino
    masses to be lighter than about $1.12$ TeV. In addition the model allows charged lepton flavor violating processes within reach of the forthcoming experiments. The scalar potential and the gauge sector of the model are analyzed and discussed in detail. Our model successfully accommodates the observed Dark matter relic density.
	
	
\end{abstract}

\maketitle



\allowdisplaybreaks
\section{Introduction}
\label{Intro}

Despite its great successes, the Standard Model (SM) does not explain the observed mass and mixing hierarchies   in the fermion sector, which remain without a
compelling explanation. It is known that in the SM, the masses of the matter fields are generated from the Yukawa interactions. In addition, the CKM  quark mixing matrix is also constructed
from the same Yukawa couplings.
To solve these puzzles, some mechanisms have been proposed. To the best of our knowledge, the first attempt to explain the huge differences in the SM fermion
masses is the Froggatt - Nielsen (FN) mechanism \cite{Froggatt:1978nt}.
According to the FM mechanism, the mass differences between generations of fermions arise from suppression factors depending on the FN charges of  the particles.
It has been noticed that in order to implement the aforementioned mechanism,
the effective Yukawa interactions have to be introduced,
thus making this theory
non-renormalizable. From this point of view, the recent mechanism proposed
by C\'arcamo, Kovalenko and Schmidt \cite{CarcamoHernandez:2016pdu} (called by CKS mechanism)
based on sequential loop suppression mechanism, is more natural since its
suppression factor arises from the loop factor $l\approx (1/4\pi)^2 $.

One of the main purposes of the models based on the gauge group $%
SU(3)_C\times SU(3)_L \times U(1)_X$ (for short, 3-3-1 model) \cite%
{Singer:1980sw, Valle:1983dk,Pisano:1991ee,Foot:1992rh,
Frampton:1992wt,Foot:1994ym,Hoang:1995vq,Hoang:1996gi} is concerned with the
search of an explanation for the number of generations of fermions.
Combined with the QCD asymptotic freedom, the 3-3-1 models provide an
explanation for the number of fermion generations. These models have nonuniversal $U(1)_X$ gauge assignments for the left handed quarks fields, thus implying that the cancellation of chiral anomalies is fulfilled when the number of $SU(3)_L$ fermionic triplets is equal to the number of $SU(3)_L$ fermionic antitriplets, which happens when the number of fermion families is a multiple of three. Some other advantages of the 3-3-1 models are: i) they solve the electric
charge quantization \cite{deSousaPires:1998jc,VanDong:2005ux}, ii) they
contain several sources of CP violation \cite{Montero:1998yw,Montero:2005yb}%
, and iii) they have a natural Peccei-Quinn symmetry, which solves the
strong-CP problem \cite{Pal:1994ba,Dias:2002gg,Dias:2003zt,Dias:2003iq}.

In the framework of the 3-3-1 models, most of the research is focused on
radiative seesaw mechanisms,
and but some involving \emph{\ nonrenormalizable} interactions introduced to explain 
the SM fermion mass and mixing pattern (see
references in Ref.\cite{CarcamoHernandez:2017cwi}).  

The FN mechanism was implemented in the 3-3-1 models in Ref.\cite%
{Huitu:2017ukq}. It is interesting to note that the FN mechanism does not produce a
new scale since the scale of the flavour breaking is the same as the
symmetry breaking scale of the model.

The CKS mechanism has been implemented for the first time in the 3-3-1 model without exotic
electric charges $(\beta = -1/\sqrt{3})$ in Ref. \cite{CarcamoHernandez:2017cwi}. The
implementation of the CKS mechanism in the 3-3-1 model leads to viable renormalizable 3-3-1 model 
that provides a dynamical explanation for the observed SM fermion mass spectrum and mixing
 parameters consistent with the SM low energy fermion flavor data \cite{CarcamoHernandez:2017cwi}. 
  It is worth mentioning that the extension of the inert 331 model of Ref. \cite{CarcamoHernandez:2017cwi} 
  contains   a residual discrete $Z^{(L_g)}_2 $ lepton
number symmetry arising from the spontaneous breaking of the global $U(1)_{L_g}$
symmetry. Under this residual symmetry,
the leptons are charged and the other particles are neutral \cite{CarcamoHernandez:2017cwi}.

However, in the mentioned work, the authors have just focused on the data
concerning fermions (both quarks and leptons including neutrino mass and
mixing), but some questions are open for the future study.

The purpose of this work is to study several phenomenological aspects of the renormalizable extension of the
 inert 3-3-1 model with sequentially loop-generated
SM fermion mass hierarchy. In particular, the constraints arising from the experimental data on 
the $\rho$ parameter, as well as those ones resulting from the charged lepton flavor violating
 process $\mu\to e\gamma$ and dark matter. Furthermore our work discusses the $Z^{\prime}$ production at proton-proton collider via quark-antiquark annihilation. To determine the oblique $\rho$ parameter constraints on the $SU(3)_L\times U(1)_X$ symmetry breaking scale $v_\chi$, which will be used to constrain the heavy $Z^{\prime}$ gauge boson mass, we proceed to study in detail the gauge and Higgs sectors of the model. In addition we determine the constraints imposed by the charged lepton flavor violating process $\mu\to e\gamma$ and dark matter on the model parameter space. In what regards the scalar potential of the model, due to the implemented symmetries, the Higgs sector is rather
simple and can be completely solved. All Goldstone bosons and the SM like
Higgs boson are defined.

The further content of this  paper is  as follows. In  Sect.~\ref{model}, we briefly present particle content and SSB of the model.
 Sect.~\ref{gboson} is devoted to gauge boson mass and  mixing. Taking into account of data on  the $\rho$ parameter, and if only contributions of the gauge bosons  are mentioned,  we will show that  the mass of the heavy neutral boson $Z'$ will be constrained  nearly to the excluded regions derived  from other experimental data such as LHC searches, and $K,D$ and $B$ meson mixing. The Higgs sector is considered in Sect. \ref{Higsector}.   The Higgs sector consists of
two parts:  the first part contains  lepton number conserving terms and the second one is lepton number violating.
 We study in details the first part and show that the Higgs
sector has all necessary ingredients. The $\rho$ parameter will be investigated including  Higgs contributions. In Sec.~\ref{LFV}, lepton flavor violating decays of the charged leptons are discussed, where  sterile neutral lepton masses are constrained. Sect. \ref{collider} is devoted to the production of the heavy $Z^\prime$ and the heavy neutral scalar
$H_4$. In Sect. \ref{DMsection}, we deal with the DM relic density. We make conclusions in Sect. \ref{conclusions}. The scalar potential of the model is given in Appendix \ref{scalarpotential}.

\section{Review of the model}
\label{model} 

To implement the CKS mechanism, only the heaviest particles such as the
exotic fermions and the top quark get masses at tree level. The next -
medium ones: bottom, charm quarks, tau and muon get masses at one-loop
level. Finally, the lightest particles: up, down, strange quarks and the
electron acquire masses at two-loop level. To forbid the usual Yukawa
interactions, the discrete symmetries should be implemented. Hence, the full
symmetry of the model under consideration is
\be
SU(3)_C \times SU( 3) _L\times U( 1)_X\times Z_4\times Z_2\times
U(1)_{L_g}\, ,  \label{eq0}
\ee
where $L_g$ is the generalized lepton number defined in Refs.~\cite%
{Chang:2006aa,CarcamoHernandez:2017cwi}. It is interesting to note that, in this model, the
light active neutrinos get their masses from a combination of linear and
inverse seesaw mechanisms at two-loop level.

As in the ordinary 3-3-1 model without exotic electric charges, the quark
sector contains the following $SU(3)_C \times SU(3)_{L}\times U(1)_{X}$
representations \cite{CarcamoHernandez:2017cwi}
\begin{align}
Q_{nL}& = \left( D_{n} \, , -U_{n} \, , J_{n} \right)^T_L\sim (
3,3^{\ast },0) ,\cm  Q_{3L}=\left( U_{3} \, , D_{3} \, , T
\right)^T_{L}\sim \left( 3,3,\fr{1}{3}\right) ,\cm  n=1,2,  \notag
\\
D_{iR}& \sim \left( 3,1,-\fr{1}{3}\right) ,\cm  U_{iR}\sim \left(
3,1,\fr{2}{3}\right),\cm  i=1,2,3,  \crn
J_{nR}&\sim \left( 3,1,-\fr{1}{3}\right) ,\cm  T_{R}\sim \left(
3,1,\fr{2}{3}\right),  \crn
\widetilde{T}_{L,R}& \sim \left( 3,1,\fr{2}{3}\right) \, , \cm
B_{L,R} \sim \left( 3,1,-\fr{1}{3}\right) \, ,  \label{eq1}
\end{align}
where $\sim$ denotes the quantum numbers for the three above subgroups,
respectively. Note that the $SU(3)_{L}$ singlet exotic up type quarks $%
\widetilde{T}_{L,R}$, down type quarks $B_{L,R}$ in the last line of Eq. (%
\ref{eq1}) are added to the quark spectrum of the ordinary 3-3-1 model in order to implement the CKS mechanism.

In the leptonic sector, besides the usual $SU(3)_L$ lepton triplets, the
model contains extra three charged leptons $E_{j(L,R)}$ ($j=1,2,3$) and four
neutral leptons, i.e, $N_{jR}$ and $\Psi _{R}$ ($j=1,2,3$). The leptonic
fields have the following $SU(3)_C \times SU(3)_{L}\times U(1)_{X}$
assignments:
\begin{align}
L_{iL}&=\left( \nu _{i} \, , e_{i} \, , \nu _{i}^{c} \right)^T_{L}\sim
\left( 1,3,-\fr{1}{3}\right) ,\cm  e_{iR}\sim (1,1,-1),\hspace*{%
1cm} i=1,2,3,  \label{L} \\
E_{1L} &\sim ( 1,1,-1) ,\cm  E_{2L}\sim (
1,1,-1) ,\cm  E_{3L}\sim ( 1,1,-1) ,  \crn
E_{1R} &\sim ( 1,1,-1) ,\cm  E_{2R}\sim (
1,1,-1) ,\cm  E_{3R}\sim ( 1,1,-1) ,  \crn
N_{1R} &\sim ( 1,1,0) ,\cm  N_{2R}\sim (
1,1,0) ,\cm  N_{3R}\sim ( 1,1,0) ,\cm
\Psi _{R}\sim ( 1,1,0) .
\end{align}
where $\nu _{iL}, \nu^{c} \equiv \nu_{R}^{c}$ and $e_{iL}$ ($e_{L},\mu
_{L},\tau _{L}$) are the neutral and charged lepton families, respectively.

The Higgs sector contains three scalar triplets: $\chi $, $\eta $ and $\rho $
and seven singlets $\va _{1}^{0}$,$\ \va _2 ^{0}$, $\xi^{0}$,$\
\phi _{1}^{+}$,$\ \phi _2 ^{+}$, $\phi _{3}^{+}$ and $\phi _{4}^{+}$.
Hence, the scalar spectrum of the model is composed of the following fields
\bea
\chi & =&\langle \chi \rangle + \chi^\prime\, \sim \left( 1,3,-\fr{1}{3}%
\right) ,  \label{eqtl1} \\
\langle \chi \rangle &= & \left( 0 \, , 0 \, , \fr{v_\chi}{\sqrt{2}}%
\right)^T\, , \hspace*{0.3cm} \chi^\prime = \left( \chi _{1}^{0} \, , \chi
_2 ^{-} \, , \fr{1}{\sqrt{2}}(R_{\chi _{3}^{0}} - i I_{\chi _{3}^{0}} ) \right)^T ,  \crn
\rho & =&\left( \rho _{1}^{+} \, , \fr{1}{\sqrt{2}}(R_{\rho }- i I_{\rho
}) \, , \rho _{3}^{+}\right)^T\sim \left( 1,3,\fr{2}{3}\right) ,  \crn
\eta & =&\langle \eta \rangle + \eta^\prime\, \sim \left( 1,3,-\fr 1 3
\right) ,  \crn
\langle \eta \rangle & = &\left(\fr{v_\eta}{\sqrt{2}} \, , 0 \, , 0
\right)^T\, , \hspace*{0.3cm} \eta^\prime =\left( \fr{1}{\sqrt{2}}(R_{\eta _{1}^{0}} -
i I_{\eta _{1}^{0}}) \, , \eta _2 ^{-} \, , \eta _{3}^{0}\right)^T,  \crn
\va _{1}^{0} & \sim& (1,1,0),\cm  \va _2 ^{0}\sim (1,1,0),
\crn
\phi _{1}^{+}& \sim & (1,1,1),\hspace*{0.5cm} \phi _2 ^{+}\sim (1,1,1),%
\hspace*{0.5cm} \phi _{3}^{+}\sim (1,1,1),\hspace*{0.5cm} \phi _{4}^{+}\sim
(1,1,1),  \crn
\xi^0& =& \langle \xi^0\rangle + \xi ^{0^\prime }\, , \langle \xi^0\rangle
= \fr{v_\xi}{\sqrt{2}}\, , \xi ^{0^\prime } = \fr{1}{\sqrt{2}}%
(R_{\xi^0} - i I_{\xi^0}) \sim (1,1,0) \, .  \label{scalarsector1}
\eea

The $Z_{4}\times Z_2 $ assignments of scalar the fields are shown in Table \ref{scalarassigments}.
\begin{table}[tbp]
\caption{Scalar assignments under $Z_{4}\times Z_2 $ }
\begin{tabular}{|c|c|c|c|c|c|c|c|c|c|c|}
\hline
& $\chi$ & $\eta$ & $\rho$ & $\va_{1}^{0}$ & $\va_2 ^{0}$ & $%
\phi_{1}^{+}$ & $\phi_2 ^{+}$ & $\phi_{3}^{+}$ & $\phi_{4}^{+}$ & $\xi^{0}$
\\ \hline
$Z_4$ & $1$ & $1$ & $-1$ & $-1$ & $i$ & $i$ & $-1$ & $-1$ & $1$ & $1$ \\
\hline
$Z_2$ & $-1$ & $-1$ & $1$ & $1$ & $1$ & $1$ & $1$ & $-1$ & $-1$ & $1$ \\
\hline
\end{tabular}%
\vspace{-0.1cm}
\label{scalarassigments}
\end{table}

The fields with nonzero lepton number are presented in Table \ref%
{nonzerolepton}.
Note that the three gauge singlet neutral leptons $N_{iR}$ as well as the elements in the third component of the
lepton triplets, namely $\nu^c_{iL}$ have lepton number equal $-1$.
\begin{table}[tbp]
\caption{Nonzero lepton number $L$ of fields}
\label{nonzerolepton}%
\resizebox{15cm}{!}{
\begin{tabular}{|c|c|c|c|c|c|c|c|c|c|c|c|c|c|c|c|c|c|}
\hline
 & $T_{L,R}$ & $J_{1L,R}$ & $J_{2L,R}$ & $\nu^c_{iL}$ & $e_{i L,R}$ & $E_{i L,R}$ & $N_{iR}$ & $\Psi_{R}$ & $\chi^0_{1}$ & $\chi^+_2 $ & $\eta^0_{3}$ & $\rho^+_{3}$ & $\phi_2^+$ & $\phi_3^+$ & $\phi_4^+$ & $\xi^0$ &$i=1,2,3$ \\
 \hline
$L$ & $-2$ & $2$ & $2$ & $-1$ & $1$ & $1$ & $-1$ & $1$ & $2$ & $2$ & $-2$ & $-2$ & $-2$ & $-2$ & $-2$ & $-2$ &  \\ \hline \end{tabular}}
\end{table}

In the model under consideration, the spontaneous symmetry breaking (SSB)
occurs by two steps \cite{CarcamoHernandez:2017cwi}. The first step is triggered by the vacuum
expectation values (VEVs) of the $\chi_3^0$ and $\xi^0$ scalar fields. At
this step, all new extra fermions, non-SM gauge bosons as well as the electrically neutral gauge
singlet lepton $\Psi_R$ gain masses. In addition, the entries of the neutral
lepton mass matrices with negative lepton number $(-1)$ also get values
proportional to $v_{\xi}$. At this step, the initial group breaks down to the direct product of the SM gauge group and the $Z_4\times Z^{(L_g)}_2 $ discrete group. The second step is triggered by $v_\eta$ providing masses for the top quark as well as for the $W$ and $Z$
gauge bosons and leaving the $SU(3)_C \times U(1) _{Q}\times Z_4\times
Z^{(L_g)}_2 $ symmetry preserved.
Here $Z^{(L_g)}_2 $ is residual symmetry where only leptons are charged, thus forbidding interactions having an odd number of leptons.
 This is crucial to guarantee the proton stability
\cite{CarcamoHernandez:2017cwi}.  Thus
\bea
&&SU(3)_C \times SU(3) _{L}\times U(1) _{X}\times Z_{4}\times Z_2 \times
U(1)_{L_g}  \crn
&{\xrightarrow{v_{\chi },v_{\xi}}}&SU(3)_C \times SU(2)
_{L}\times U(1) _{Y}\times Z_4\times Z^{(L_g)}_2   \crn
&&{\xrightarrow{v_{\eta }}}SU(3)_C \times U(1) _{Q}\times Z_{4}\times
Z^{(L_g)}_2  .  \label{ssb}
\eea
A consequence of the chain in (\ref{ssb}) is
\be
v_{\eta }=v=246\mbox{GeV}\ll v_{\chi }\sim v_{\xi}\sim \mathcal{O}(10)\, %
\mbox{TeV}.  \label{eq2}
\ee

The corresponding Majoron associated to the spontaneous breaking of the $U(1)_{L_g}$ global symmetry is a gauge-singlet scalar and, therefore, unobservable.

An explanation for the relation $v_{\chi }\sim v_{\xi}\sim \mathcal{O}(10)$ \mbox{TeV} is provided in the following. The present lower limits on the $Z^{\prime }$ gauge boson mass in $3%
\text{-}3\text{-}1$ models arising from LHC searches, reach around $2.5$
\text{TeV} \cite{Salazar:2015gxa}. These bounds can be translated into
limits of about 6.3 TeV on the $SU(3)_{C}\times SU\left( 3\right) _{L}\times
U\left( 1\right) _{X}$ gauge symmetry breaking scale $v_{\chi }$.
Furthermore, electroweak data from the decays $B_{s,d}\rightarrow \mu
^{+}\mu ^{-}$ and $B_{d}\rightarrow K^{\ast }(K)\mu ^{+}\mu ^{-}$ set lower
bounds on the $Z^{\prime }$ gauge boson mass ranging from $1$ TeV up to $3$
TeV \cite%
{CarcamoHernandez:2005ka,Martinez:2008jj,Buras:2013dea,Buras:2014yna,Buras:2012dp}. Furthermore, as shown in Ref. \cite{Huyen:2012uk}, the experimental data on $K$, $D$ and $B$ meson mixings set a lower bound of about $4$ TeV for the $Z^{\prime }$ gauge boson mass in 3-3-1 models, which translates in a lower limit of about $10$ TeV for the $SU\left( 3\right) _{L}\times
U\left( 1\right) _{X}$ gauge symmetry breaking scale $v_{\chi }$.

Finally, to close this section we provide a justification of the role of the different particles of our model:
\begin{enumerate}
  \item The presence of the $SU(3)_{L}$ scalar singlet $\phi
_{3}^{+}$, is needed to generate two loop level down and strange
quark masses, as shown in Ref.~\cite{CarcamoHernandez:2017cwi}. Besides that, in order
to implement a two loop level radiative seesaw mechanism for the generation
of the up, down and strange quark masses as well as the electron mass, the $%
Z_{4}$ charged $SU(3)_{L}$ scalar singlets $\varphi _{1}^{0}$,$\ \varphi
_{2}^{0}$, $\phi _{1}^{+}$,$\ \phi _{2}^{+}$ (which do not acquire a vacuum
expectation value) are also required in the scalar sector. The $Z_{4}$
charged $SU(3)_{L}$ scalar singlet $\varphi _{1}^{0}$ is also needed to generate one loop level masses for the charm and bottom quarks as well as for the tau and muon
leptons. The $Z_{4}$ charged $%
SU(3)_{L}$ scalar singlets $\varphi _{2}^{0}$ and $\phi _{3}^{+}$ as well as
the $SU(3)_{L}$ scalar singlet $\phi _{4}^{+}$, neutral under $Z_{4}$ are
also crucial for the implementation of two loop level linear and inverse
seesaw mechanisms that give rise to the light active neutrino masses. The $SU(3)_{L}$ scalar
singlet $\xi^{0}$ is introduced to spontaneously break the $U(1)_{L_g}$ generalized lepton number symmetry and thus giving rise to a tree-level mass for the right handed Majorana neutrino $\Psi_R$. It is crucial for generating two loop-level masses for the down and strange quarks. 
\item The $SU(3)_{L}$ singlet exotic down type quark,
i.e. $B$, is crucial for the implementation of the one loop level radiative
seesaw mechanism that generate the bottom quark mass. The $SU(3)_{L}$
singlet exotic up type quarks, i.e., $\widetilde{T}_{1}$ and $\widetilde{T}%
_{2}$, are needed to generate a one loop level charm quark mass as well as
two loop level down and strange quark masses. The three $SU(3)_{L}$ singlet
exotic charged leptons, i.e., $E_{j}$ ($j=1,2,3$), are required in order to
provide the radiative seesaw mechanisms that generate one loop level tau
and muon masses and two loop level electron mass. The four right handed
Majorana neutrinos, i.e., $N_{jR}$ ($j=1,2,3$), $\Psi _{R}$, are crucial for
the implementation of the two loop level linear and inverse seesaw
mechanisms that give rise to the light active neutrino masses.
\end{enumerate}

\section{Gauge bosons}
\label{gboson}

\subsection{Gauge boson masses and mixing}
\label{gmass}

After SSB, the gauge bosons get masses arising from the kinetic terms for
the $\eta$ and $\chi$ $SU(3)_L$ scalar triplets, as follows:
\be
L_{mass}^{gauge} = (D_\mu \langle \chi \rangle)^\dag D^\mu \langle \chi
\rangle + (D_\mu \langle \eta \rangle)^\dag D^\mu \langle \eta \rangle\, ,
\label{eq181}
\ee
with the covariant derivative for triplet defined as
\be
D_\mu = \partial_\mu - i g A_{\mu a} \fr{\la_a}{2} - i g_X X \fr{%
\la_9}{2}B_\mu \, ,  \label{eq182}
\ee
where $g$ and $g_X$ are the gauge coupling constants of the $SU(3)_L$ and $%
U(1)_X$ groups, respectively. Here, $\la_9 = \sqrt{2/3}\text{ diag}
(1,1,1)$ is defined such that $\text{Tr} (\la_9\la_9) =2$, similarly
as the usual Gell-Mann matrix $\la_a, a= 1, 2, 3, \cdots ,8$. By
matching gauge the coupling constants at the $SU(3) _{L}\times U(1) _{X}$
symmetry breaking scale, the following relation is obtained 
\cite{Hoang:1995vq}
\be
t \equiv \fr{g_X}{g} = \fr{3\sqrt{2} \sin \theta_W(M_{Z^\prime })}{%
\sqrt{3-4\sin^2 \theta_W(M_{Z^\prime })}}\, .  \label{eq183}
\ee

Let us provide the definition of the Weinberg angle $\theta_W$. As in the
SM, one puts $g^\prime = g \tan \theta_W$, where $g^\prime $ is gauge
coupling of the $U(1)_Y$ subgroup satisfying the relation~\cite{Hoang:1995vq}
\be
g^\prime = \fr{\sqrt{3} g g_X}{\sqrt{18 g^2 - g_X^2}} \, .
\label{eq1810}
\ee
Thus
\be
\tan \theta_W = \fr{\sqrt{3} g_X}{\sqrt{18 g^2 - g_X^2}} \, .
\label{eq1811}
\ee
Denoting
\be
W_\mu^\pm = \fr{1}{\sqrt{2}}\left( A_{\mu 1} \mp i A_{\mu 2} \right)\, ,
\hspace*{0.3cm} Y_\mu^\pm = \fr{1}{\sqrt{2}}\left( A_{\mu 6} \pm i A_{\mu
7} \right)\, , \hspace*{0.3cm} X_\mu^0 = \fr{1}{\sqrt{2}}\left( A_{\mu 4}
- i A_{\mu 5} \right)\, ,  \label{eq184}
\ee
and substituting (\ref{eq182}) and (\ref{eq184}) into (\ref{eq181}) one gets the following squared masses for the charged/non-Hermitian gauge bosons
\be
m^2_W = \fr{g^2}{4}v_\eta^2 \, , \hspace*{0.3cm} M^2_{X^0} = \fr{g^2}{4}%
\left(v^2_\chi +v_\eta^2\right) \, , \hspace*{0.3cm} M^2_{Y} = \fr{g^2}{4}%
v_\chi^2\, ,  \label{eq185}
\ee
where $v_\eta = v =246 $ GeV, as expected.

From Eq.(\ref{eq185}) we find the following gauge boson mass squared splitting
\be
M^2_{X^0} - M^2_Y = m^2_W \, .  \label{eq186}
\ee

For neutral gauge bosons, the squared mass mixing matrix has the form
\be
L_{mass}^{ngauge} = \fr 1 2 V^T M^2_{ngauge} V\, ,  \label{eq187}
\ee
where $V^T = (A_{\mu 3}, A_{\mu 8}, B_\mu)$ and
\be
M^2_{ngauge} =\fr{g^2}{4}\left(
\begin{array}{ccc}
v_\eta^2 & \fr{v_\eta^2}{\sqrt{3}} & -\fr{2 t}{3\sqrt{6}}v_\eta^2 \\
& \fr 1 3 (4 v^2_\chi + v^2_\eta) & \fr{2 t}{9\sqrt{2}}(2 v^2_\chi -
v_\eta^2) \\
&  & \fr{2 t^2}{27}(v^2_\chi + v_\eta^2)%
\end{array}
\right)\, .  \label{eq188}
\ee
The down-left entries in (\ref{eq188}) are not written, due to the fact that
the above matrix is symmetric.

The matrix in (\ref{eq188}) has vanishing determinant, thus giving rise to a massless gauge boson, which corresponds to the photon.
The diagonalization of the squared mass matrix for neutral gauge bosons of Eq. (\ref{eq188}) is divided in two steps. In the
first step, the massive fields are identified as
\bea
A_\mu & = & s_W A_{\mu 3} + c_W \left(-\fr{t_W}{\sqrt{3}} A_{\mu 8} +
\sqrt{1-\fr{t^2_W}{3}}B_\mu \right)\, ,  \crn
Z_\mu & = & c_W A_{\mu 3} - s_W \left(-\fr{t_W}{\sqrt{3}} A_{\mu 8} +
\sqrt{1-\fr{t^2_W}{3}}B_\mu \right)\, ,  \label{eq189} \\
Z^\prime_\mu & = & \sqrt{1-\fr{t^2_W}{3}} A_{\mu 8} + \fr{t_W}{\sqrt{3}}
B_\mu \, ,  \nn
\eea
where we have denoted $s_W = \sin \theta_W, \, c_W = \cos \theta_W\, ,\, t_W
= \tan \theta_W $.  The coupling of the photon $A_{\mu}$ gives $e = g s_W $. After the first step, the squared mass matrix is a block
diagonal one  in the new basis $(A_{\mu}, Z_{\mu},Z'_{\mu})$, where the entry in the top is zero (due to the masslessness of
the photon), while the $2 \times 2$ matrix for $(Z_\mu, Z^\prime_\mu)$ in
the bottom has the form
\be
M^2_{(2\times 2)} = \left(
\begin{array}{cc}
M_Z^2 & M^2_{Z Z^\prime } \\
M^2_{Z Z^\prime } & M^2_{Z^\prime }%
\end{array}
\right)\, .  \label{eq1812}
\ee
The matrix elements in (\ref{eq1812}) are given by
\bea
M_Z^2 & = & \fr{g^2 v^2_\eta}{4 c^2_W} = \fr{m^2_W}{c^2_W}\, ,
\label{eq1813} \\
M^2_{ZZ^\prime } & = & \fr{g^2}{4 c^2_W\sqrt{3-4s^2_W}}%
v^2_\eta(1-2s_W^2)\, ,  \crn
M_{Z^\prime }^2 & = & \fr{g^2c^2_W}{4 (3-4s^2_W)}\left[ 4 v^2_\chi +
\fr{v^2_\eta (1-2 s_W^2)^2}{c^4_W} \right] \, .  \notag
\eea
Note that our formula of $M^2_{Z^\prime }$ is consistent with that given
in \cite{Buras:2012dp}.

In the last step of diagonalization, the $Z-Z'$ mixing angle $\phi$ and  mass eigenstates  $Z^{1,2}$ are determined as
\bea
\tan 2 \phi &=& \fr{2 M^2_{ZZ^\prime } }{M_{Z^\prime }^2 - M_{Z}^2}\,, 
\label{eq1815}\\
Z^1_{\mu } & = & Z_\mu \cos \phi - Z^\prime_{\mu} \sin \phi \, ,  \crn
Z^2_{\mu } & = & Z_\mu \sin \phi + Z^\prime_{\mu} \cos \phi \, .
\label{eq1814}
\eea
Our definition of $\phi$ is consistent with
that in Ref.~\cite{Hoang:1999yv} needed to study the $\rho$ parameter.

The masses of physical neutral gauge bosons are determined as
\bea
M^2_{Z^1} & = & \fr 1 2 \left\{ M_{Z^\prime }^2 + M_{Z}^2 - \left[%
(M_{Z^\prime }^2 - M_{Z}^2)^2 + 4 (M^2_{ZZ^\prime })^2\right]^{\fr 1
2} \right\}\, ,  \crn
M^2_{Z^2} & = & \fr 1 2 \left\{ M_{Z^\prime }^2 + M_{Z}^2 + \left[%
(M_{Z^\prime }^2 - M_{Z}^2)^2 + 4 (M^2_{ZZ^\prime })^2\right]^{\fr 1
2} \right\}\, .  \label{eq1816}
\eea

In the limit $v^2_\chi \gg v^2_\eta$, one approximates
\bea
M^2_{Z^1} & \simeq & M_{Z}^2 -\fr{ (M^2_{ZZ^\prime })^2 }{M_{Z^{\prime
}}^2} + M_{Z}^2\times \mathcal{O}\left( \fr{v_\eta^4}{v_\chi^4}\right)\, ,
\label{eq212} \\
M^2_{Z^2} & \simeq & M_{Z^\prime }^2 +\fr{ (M^2_{ZZ^\prime })^2}{%
M_{Z^\prime }^2} +M_{Z}^2\times \mathcal{O}\left( \fr{v_\eta^4}{v_\chi^4}%
\right) \simeq M_{Z^\prime }^2 \, .  \label{eq213}\\ 
\tan \phi &\simeq&  \fr{(1-2 s_W^2)\sqrt{3-4s^2_W}}{4 c^4_W}\left( \fr{%
v_\eta^2}{v_\chi^2} \right)\, .  \label{eq1817}
\eea

\subsection{Limit on $Z^\prime $ mass from the $\protect\rho$ parameter}
\label{rho}

The presence of the non SM particles modifies the oblique
corrections of the SM, the values of which have been extracted from high
precision experiments. Consequently, the validity of our model depends on
the condition that  the non SM particles do not contradict those
experimental results. Let us note that one of the most important observables
in the SM is the $\rho$ parameter defined as
\be
\rho =\fr{m^2_W}{c_W^2 M^2_{Z}}\, .  \label{eq261}
\ee

For the model under consideration, one-loop contributions of the new heavy gauge bosons to  the oblique correction lead to the
following form of the $\rho$ parameter \cite{Hoang:1999yv}
\bea
\rho - 1 &\simeq & \tan^2 \phi \left( \fr{ M^2_{Z^\prime }}{m^2_{Z }} -
1 \right) + \fr{3\sqrt{2}G_F}{16\pi^2} \left[ M_+^2 + M_0^2 + \fr{%
2M_+^2M_0^2}{M_+^2-M_0^2} \ln \fr{M_0^2}{M_+^2} \right]  \crn
& & - \fr{\al(m_Z)}{4\pi\ s^2_W} \left[\ t^2_W \ln \fr{M_0^2}{M_+^2}
+ \fr{\varep^2(M_+,M_0)}{2} + O( \varep^3(M_+,M_0)) \right] \, ,
\label{eq822}
\eea
where $M_0 = M_{X^0}, M_+ = M_{Y^+}$ and $\varep(M,m) \equiv \fr{M^2
- m^2}{m^2}$.

Combining with Eq.~(\ref{eq186}), one gets
\bea
\rho - 1 &\simeq & \tan^2 \phi \left( \fr{ M^2_{Z^\prime }}{m^2_{Z }} -
1 \right) + \fr{3\sqrt{2}G_F}{16\pi^2} \left[ 2 M_{Y^+}^2 + m^2_W - \fr{%
2M_{Y^+}^2 (M_{Y^+}^2 +m^2_W) }{m^2_W} \ln \fr{(M_{Y^+}^2 +m^2_W)}{%
M_{Y^+}^2} \right]  \crn
& & - \fr{\al(m_Z)}{4\pi\ s^2_W} \left[\ t^2_W \ln \fr{(M_{Y^+}^2
+m^2_W)}{M_{Y^+}^2} + \fr{m_W^4}{2(M_{Y^+}^2 +m^2_W)^2} \right] \, ,
\label{eq822}
\eea
where $\al (m_Z) \approx \fr{1}{128}$%
~\cite{Tanabashi:2018oca}.

\begin{figure}[ht]
\centering
\begin{tabular}{cc}
\includegraphics[width=7.8cm]{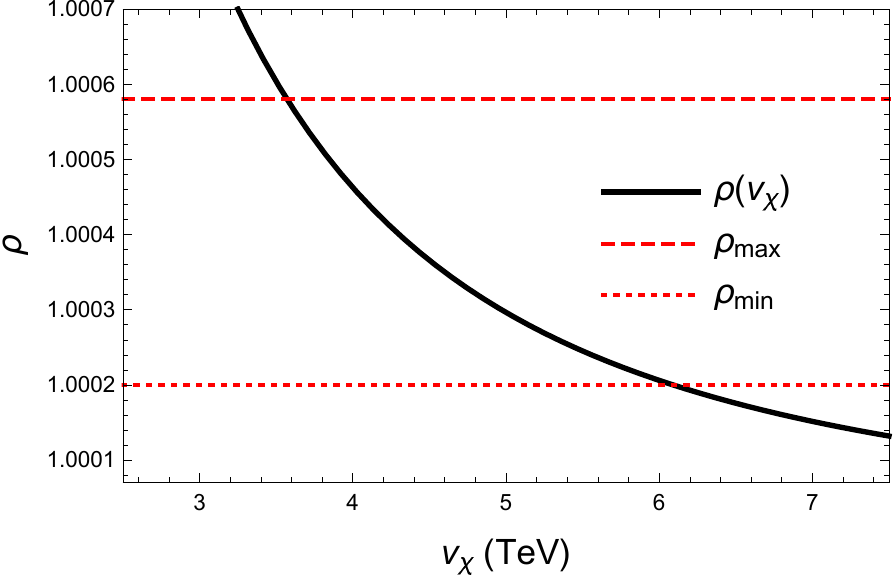} & %
\includegraphics[width=7.8cm]{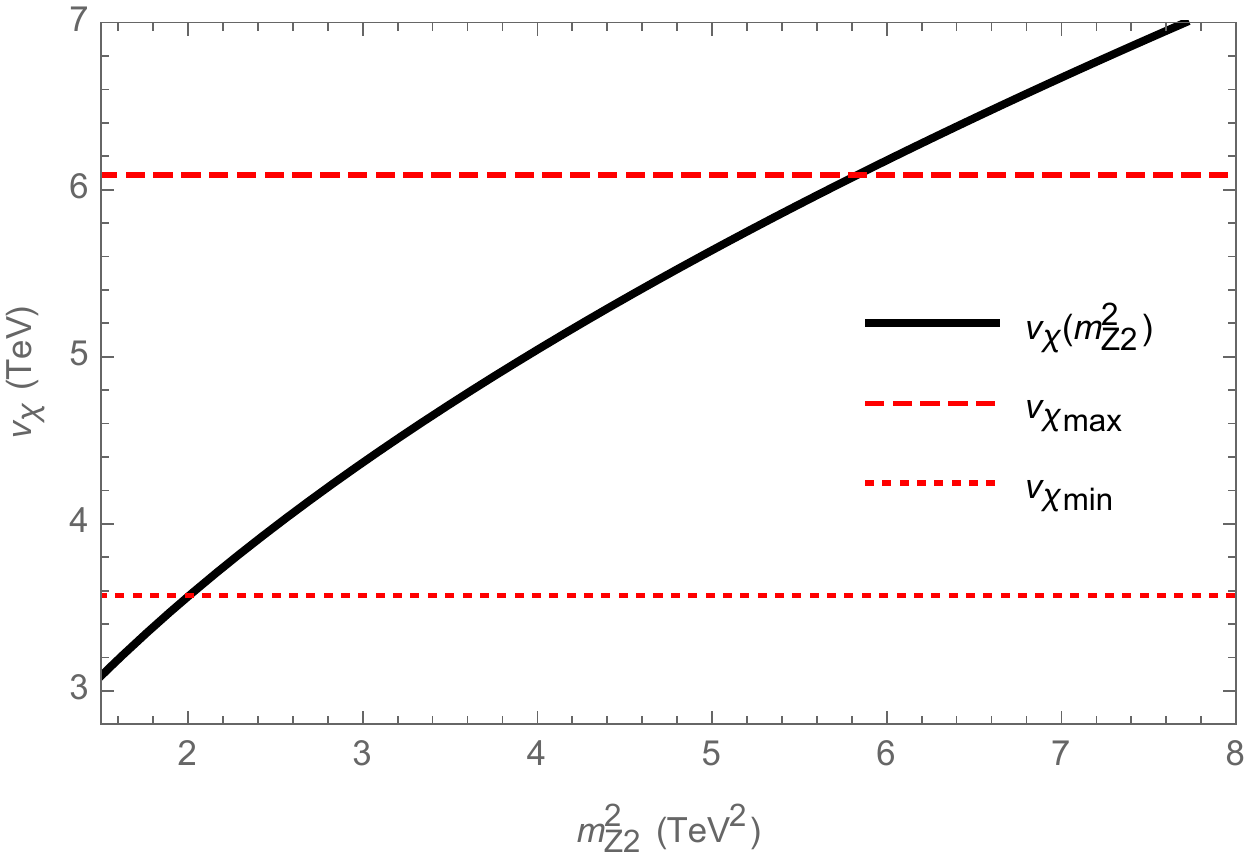} \\
&
\end{tabular}%
\caption{Left-panel: $\rho$ parameter as a function of $v_\protect\chi$, upper and a lower horizontal lines are an upper a lower limits in (\protect
	\ref{eq262t}) . Right-panel:  Relation between $v_\protect\chi $ and $M^2_{Z_2}$,  upper horizontal lines are an upper and a lower limits of  $v_\protect\chi$, respectively.}
\label{fig1}
\end{figure}

Taking into account $s^2_W = 0.23122$ \cite{Tanabashi:2018oca} and
\bea
\rho &= & 1.00039 \pm 0.00019\, ,  \label{eq262t}
\eea
we have plotted $\De \rho $ as a function of $v_\chi$ in Fig. \ref{fig1}
(the left-panel).
From figure \ref{fig1} (the left-panel), it follows
\be
3.57\, \mbox{TeV}\, \leq v_\chi \, \leq 6.09\, \text{TeV}.  \label{eq851}
\ee

Substituting (\ref{eq851}) into (\ref{eq213}) and evaluating in figure \ref%
{fig1}(the right-panel)
we get a bound on the $Z^\prime $ mass as follows
\be
1.42\, \mbox{TeV}\, \leq M_{Z^2}\, \leq \, 2.42 \, \mbox{TeV} \, .
\label{eq852}
\ee

Then, the bilepton gauge boson mass is constrained to be in the range:
\be
465\, \mbox{GeV}\, \leq M_{Y}\, \leq \, 960 \, \mbox{GeV} \, ,  \label{eq853}
\ee
where  $m_W = 80.379 $ GeV~\cite{Tanabashi:2018oca}. The above limit is stronger than the one obtained from the wrong muon decay $M_Y \geq 230$  GeV \cite{Dong:2006mg}.

It is worth mentioning that the second term in (\ref{eq822}) is much smaller
the first one. Consequently, the limit derived from the tree level contribution
is very close to the one obtained when we consider the radiative corrections arising from heavy vector exchange.

From LHC searches, it follows that the lower bound on the $Z^\prime $
boson mass in 3-3-1 models  ranges from  $2.5$ to $3$ TeV \cite{Salazar:2015gxa,Queiroz:2016gif}.
From the decays $B_{s,d}\rightarrow \mu ^{+}\mu ^{-}$ and $B_{d}\rightarrow K^{\ast }\, (K)\,
	\mu ^{+}\mu ^{-}$ \cite{CarcamoHernandez:2005ka,Martinez:2008jj,Buras:2013dea,Buras:2014yna,Buras:2012dp},  the lower limit on the $Z^\prime $ boson mass ranges from $1$ TeV to $3$ TeV.    We will show that when scalar contributions to the $\rho$ parameter are included, there will  exist  allowed  $m_{Z_2}$ values  larger than the range given in \eqref{eq852}, so that they satisfy  the recent lower bounds concerned from LHC searches.

For conventional notation, hereafter we will call $Z^1$ and $Z^2$ by $Z$ and
$Z^\prime$, respectively.

Now we turn into the main subject - the Higgs sector.

\section{Analysis of the lepton number conserving part of the scalar potential}
\label{Higsector}

Below we present lepton number conserving part $V_{LNC}$ of the scalar potential of the model shown in Appendix \ref{scalarpotential}. Expanding the
Higgs potential around theVEVs, one gets the scalar potential minimization conditions at tree level as follows
\bea
w_3 & = & 0 \, ,  \label{conditionn} \\
- \mu_\chi^2 & = & v_\chi^2 \la_{13}+ \fr{1}{2} v_\eta^2
\la_{5}+ \fr{1}{2}\la_{\chi\xi} v_\xi^2 \, ,  \crn
- \mu_{\eta}^2 & =& v_\eta^2  \la_{17}+ \fr{1}{2}%
v_\chi^2 \la_{5}+ \fr{1}{2}\la_{\eta\xi} v_\xi^2 \, ,
\label{conditions} \\
- \mu_{\xi}^2  &=& \fr{1}{2}\la_{\chi\xi} v_\chi^2 + \fr{1}{2}%
\la_{\eta\xi} v_\eta^2 +\la_\xi v_\xi^2\, .  \nn
\eea

From the analysis of the scalar potential, taking into account 
the constraint conditions of Eq.(\ref{conditionn}), we find that the charged scalar
sector is composed of two massless fields, i.e., $\eta_2^+$ and $\chi_2^+$ which are
the Goldstone bosons eaten by the longitudinal components of the $W^+$ and $Y^+$ gauge bosons, respectively.
The other massive electrically charged fields are $\phi_1^+, \phi_2^+ $ and $\phi_4^+$ whose masses are respectively given by:
\bea
m^2_{\phi_1^+} &=& \mu^2_{\phi_1^+}  +\fr{1}{2}\left[v_\chi^2
\la^{\chi\phi}_{1} +v_\eta^2 \la^{\eta\phi}_{1} +v_\xi^2
\la^{\phi\xi}_{1}\right] \, ,  \crn
m^2_{\phi_2^+} &=& \mu^2_{\phi_2^+}  +\fr{1}{2}\left[v_\chi^2
\la^{\chi\phi}_2  +v_\eta^2 \la^{\eta\phi}_2  +v_\xi^2
\la^{\phi\xi}_2 \right]\, ,  \label{chiggsm} \\
m^2_{\phi_4^+} &=& \mu^2_{\phi_4^+}  +\fr{1}{2}\left[v_\chi^2
\la^{\chi\phi}_{4} +v_\eta^2 \la^{\eta\phi}_{4} +v_\xi^2
\la^{\phi\xi}_{4}\right]\, .  \nn
\eea
In addition, the basis ($\rho_1^+\, ,\,\rho_3^+\, ,\,\phi_3^+$)  corresponds to 
	the following squared mass  matrix 
\bea
M^2_{charged} & = & \left(
\begin{matrix}
A + \fr 1 2 v_\eta^2 \left(\la_{6} \!+ \!\!\la_{9}\right) & 0 &
\fr{1}{2}v_\eta v_\xi \la_3 \\
0 & A + \fr 1 2 \left(v_\chi^2 \la_{7} + v_\eta^2 \la_{6}\right) &
\fr{1}{\sqrt{2}} v_\chi w_2 \\
\fr{1}{2}v_\eta v_\xi \la_3 & \fr{1}{\sqrt{2}} v_\chi w_2 &  \mu^2_{\phi_3^+}  + B_3
\end{matrix}
\right)\, ,  \label{eq761}
\eea
where we have used the following notations
\bea
A &\equiv &\mu_{\rho}^2  + \fr{1}{2}\!\left[v_\chi^2 \la_{18} +
\la_{\rho\xi}v^2_\xi\right] \, , \quad 
B_i  \equiv \fr{1}{2}\left(v_\chi^2 \la^{\chi\phi}_{i} +v_\eta^2
\la^{\eta\phi}_{i} +v_\xi^2 \la^{\phi\xi}_{i}\right)\, , \hspace*{%
0.3cm} i= 1, 2, 3, 4 \, .  \label{chiggsm}
\eea
From (\ref{eq761}), it follows that in the limit $v_\eta \ll v_\xi$, $%
\rho_1^+$ is a physical field with mass
\be
m^2_{\rho_1^+} = A + \fr 1 2 v_\eta^2 \left(\la_{6} + \la_{9}\right) \, ,  \label{eq9101}
\ee
 while the  two massive \textit{bilepton} scalars $\rho_3^+$ and $\phi_3^+$ mix with  each
other.

Now we turn into CP-odd Higgs sector. There are three massless fields: $%
I_{\chi_3^0}\,, I_{\eta_1^0}$ and $I_{\xi^0}$.
 The field $I_{\va_2}$ is a physical state itself with 
  squared mass
\be
m^2_{I_{\va_2}} = \mu^2_{\va_2}+ B^\prime_2\, ,  \label{mcp1}
\ee
where
\be
B^\prime_n \equiv \fr{1}{2}\left(v_\chi^2 \la^{\chi\va}_{n}
+v_\eta^2 \la^{\eta\va}_{n} +v_\xi^2
\la^{\va\xi}_{n}\right)\, , \hspace*{0.3cm} n= 1, 2\, .
\label{mcp1t}
\ee
The squared mass matrix of  the four remaining CP-odd Higgs fields  separate into two  block diagonal submatrices  corresponding  to the two original base  $( I_{\chi_1^0}\, , I_{\eta_3^0})$ and $(I_{\va_1}\, , I_{\rho})$ , namely 
\be
m^2_{CPodd1} = \fr{\la_8}{2} \left(
\begin{array}{cc}
	v_\eta^2 & - v_\chi v_\eta \\
	- v_\chi v_\eta & v_\chi^2%
\end{array}
\right),  \, \;\;  m^2_{CPodd2} = \left(
\begin{array}{cc}
	\mu^2_{\va_1} + B^\prime_1 -C\; & \fr{1}{2}v_\chi v_\eta
	(\la_{1}-\la_2 ) \\
	\fr{1}{2}v_\chi v_\eta (\la_{1}-\la_2 ) & A +\fr{ \la_6}{2}%
	v_\eta^2%
\end{array}
\right)\,,  \label{cpod1}
\ee
where 
 \be
 C \equiv  v^2_\chi \la_{22} + v^2_\eta \la_{24} +
 v^2_\xi\la_{25} \,  \label{eq763}
 \ee 
 The first  matrix in  (\ref{cpod1}) provides two mass eigenstates, where one of them are massless,
 \bea
 G_1 &= & \cos \theta_a I_{\chi_1^0}+ \sin \theta_a I_{\eta_3^0} \, , \quad m_{G_1}=0, \crn
 A_1 &= &- \sin \theta_a I_{\chi_1^0} + \cos \theta_a I_{\eta_3^0} \, , \quad m^2_{A_1} = \fr{\la_8v^2_\chi}{2 \cos^2\theta_a } \, ,
 \label{eq752}
 \eea
 where
 \be
 \tan \theta_a = \fr{v_\eta}{v_\chi} \, .  \label{eq753}
 \ee
The physical states relating to the second  matrix in  (\ref{cpod1}) are
\be
\left(
\begin{array}{c}
	A_2 \\
	A_3 \\
\end{array}
\right) = \left(
\begin{array}{cc}
	\cos \theta_\rho & \sin \theta_\rho \\
	- \sin \theta_\rho & \cos \theta_\rho%
\end{array}
\right) \left(
\begin{array}{c}
	I_{\va_1} \\
	I_\rho \\
\end{array}
\right)\, ,  \label{eq765}
\ee
where the mixing angle is given by
\be
\tan 2\theta_\rho = \fr{v_\chi v_\eta (\la_{1}-\la_2 ) }{%
	\left(\mu^2_{\va_1} - C+B'_1 - A - \fr{ \la_6}{2}v_\eta^2\right)}
\, .  \label{eq766}
\ee
Their squared masses  are 
\be
m^2_{A_{2,3}} =  \fr{1}{2}\left\{A+D_1 \mp \sqrt{(A-D_1 )^2+v^2_\eta\left[
	2(A-D_1) \la_6 +v^2_\eta \la^2_6+v^2_\chi
	(\la_{13}-\la_{14})^2\right]} \right\}\, , 
\label{eq-noname-01}
\ee 
where $D_1 =\mu^2_{\va_1} +B^\prime_1 - C +\fr 1 2 v_\eta^2 \la_6\;$.

Next, the CP-even scalar sector is our task.   We find that 
  $R_{\va_2}$ is physical  with mass
\be
m^2_{R_{\va_2}}=m^2_{I_{\va_2}} 
= \mu^2_{\va_2} + \fr{1}{2}\left(v_\chi^2
\la^{\chi\va}_2  +v_\eta^2 \la^{\eta\va}_2  +v_\xi^2
\la^{\va\xi}_2 \right) \, .  \label{eq751}
\ee 

As mentioned in Ref.~\cite{CarcamoHernandez:2017cwi},  the lightest scalar $\va_2^0$ is a
possible DM candidate   with light mass  smaller than 1 TeV.  Therefore, Eq.~(\ref{eq751}) 
suggests a  reasonable assumption 
\be
\mu^2_{\va_2} = - \fr 1 2 \left(v_\chi^2 \la^{\chi\va}_2
+v_\xi^2 \la^{\va\xi}_2 \right)\,.   \label{eq9191}
\ee
In this case, the model contains the complex scalar DM $\va_2^0$ with
mass $m^2_{R_{\va_2}}=m^2_{I_{\va_2}} = \fr 1 2  \la^{\eta\va}_2 v_\eta^2$. 

There are  other seven CP-even Higgs components which the squared mass matrix separates  into  two $2\times2$ and one $3\times 3$ independent matrices.  The $2\times 2$ matrices  are 
\be
m^2_{CPeven1} = \fr{\la_8}{2} \left(
\begin{array}{cc}
	v_\eta^2 & v_\chi v_\eta \\
	v_\chi v_\eta & v_\chi^2%
\end{array}
\right), \;\;   m^2_{CPeven2} = \left(
\begin{array}{cc}
	A + \fr{\la_6}{2}v^2_\eta & -\fr{1}{2}v_\chi v_\eta
	(\la_{1}+\la_2 ) \\
	-\fr{1}{2}v_\chi v_\eta (\la_{1}+\la_2 ) & \mu^2_{\va_1} + C
	+ B^\prime_1
\end{array}
\right)\,,  \label{cpeven1}
\ee	
corresponding to the two original base $( R_{\chi_1^0}\, , R_{\eta_3^0})$ and $(R_\rho\, , R_{\va_1})$, 
respectively.   The physical states of the first matrix in \eqref{cpeven1} are determined as follows,
\bea
R_{G_1} &= & \cos \theta_a R_{\chi_1^0} + \sin \theta_a R_{\eta_3^0} \, , \quad m_{R_{G_1}}=0,
\crn
H_1 &= & - \sin \theta_a R_{\chi_1^0} + \cos \theta_a R_{\eta_3^0} \, ,\quad m^2_{H_1} = m^2_{A_1} = \fr{\la_8v^2_\chi}{2 \cos^2\theta_a }.  \, 
\label{eq754}
\eea
The physical states of the second matrix in  (\ref{cpeven1}) are
\be
\left(
\begin{array}{c}
	H_2 \\
	H_3 \\
\end{array}
\right) = \left(
\begin{array}{cc}
	\cos \theta_r & \sin \theta_r \\
	- \sin \theta_r & \cos \theta_r%
\end{array}
\right) \left(
\begin{array}{c}
	R_{\rho} \\
	R_{\va_1} \\
\end{array}
\right)\, ,  \label{eq767}
\ee
where the mixing angle is 
\be
\tan 2\theta_r = \fr{v_\chi v_\eta (\la_{1}+\la_2 ) }{%
	\left(\mu^2_{\va_1} + C+ B^\prime_1 - A - \fr{ \la_6}{2}v_\eta^2\right)}
\,   \label{eq766}
\ee
and their squared masses are 
\be  \label{eq-noname-01}
m^2_{H_{2,3}} =  \fr{1}{2}\left\{A+D_2 \mp \sqrt{(A-D_2 )^2 +v^2_\eta\left[
	2(A-D_2) \la_6 +v^2_\eta \la^2_6+v^2_\chi
	(\la_{13}+\la_{14})^2\right]} \right\}\, , 
\ee 
where $D_2 =\mu^2_{\va_1} + B^\prime_1 + C +\fr 1 2 v_\eta^2 \la_6\;.$

The  squared mass matrix corresponding to the basis $(R_{\chi_3^0}\, , R_{\eta_1^0}\, , R_{\xi^0})$ is 
\be
m^2_{CPeven3} = \left(
\begin{array}{ccc}
	2 v_\chi^2 \la_{13} & v_\chi v_\eta \la_5 & \la_{\chi\xi} v_\chi
	v_\xi \\
	v_\chi v_\eta \la_5 & 2 v^2_\eta \la_{17} & \la_{\eta\xi} v_\eta
	v_\xi \\
	\la_{\chi\xi} v_\chi v_\xi & \la_{\eta\xi} v_\eta v_\xi &
	2\la_\xi v^2_\xi%
\end{array}
\right) \, ,  \label{cpeven3}
\ee 
which contains a SM-like Higgs boson found by LHC. The mass eigenstates will  be  discussed using simplified conditions.

Let us summarize the Higgs content:
\ben
\item In the charged scalar sector: there are two Goldstone bosons $\eta^-$
and $\chi^-$ eaten by the gauge bosons $W^-$ and $Y^-$. Three massive
charged Higgs bosons are $\phi_1^+\, , \phi_2^+ $ and $\phi_4^+$. The
remaining fields $\rho_1^+\, ,\,\phi_3^+ $ and $\rho_3^+$ are mixing.

\item In the CP-odd scalar sector: there is one massless Majoron scalar $I_{\xi^0}$ which is denoted by $G_M$. Fortunately, it is a gauge singlet,
therefore, is \emph{phenomenologically harmless}. Two massless scalars $
I_{\eta_1^0}$ and $I_{\chi_3^0}$ are Goldstone bosons for the gauge bosons $Z$ and
$Z^\prime$, respectively. There exists another massless state denoted by $%
G_1 $, its role will be discussed below. Here we just mention that in the
limit $v_\eta \ll v_\chi$, this field is $I_{\chi_1^0}$. The massive CP-odd
field are $I_{\va_2}$, $A_1$ and other two $I_{\va_1}\, , I_{\rho}$
are mixing.

\item In the CP-even scalar sector: There is one massless field: $R_{G_2}$,
and in the limit $v_\eta \ll v_\chi$, it tends to $R_{\chi_1^0}$.
Combination of $G_1$ and $R_{G_1}$ is Goldstone boson for neutral bilepton
gauge boson $X^0$, namely $G_{X^0} = \fr{1}{\sqrt{2}}(R_{G_1} - i G_1)$.
The massive fields are: $R_{\va_2}, H_1, H_2$ and three massive $R_\chi\, , R_{\eta}\, , R_{\xi^0}$ and the SM-like Higgs boson $h$. Note
that \emph{there exists degeneracy in Eqs.~\eqref{eq751} and \eqref{eq754} when the contribution arising from $Z_2\times Z_4$ soft breaking scalar interactions is not considered.} 
Thus, the complex scalar $\va_2$ has mass given by Eq.~\eqref{eq751}, which  is consistent with the prediction in Ref.~\cite{CarcamoHernandez:2017cwi}. To be a DM candidate, the  condition \eqref{eq9191} can be used to eliminate  the terms with large VEVs such as $v_\chi$ and $v_\xi$. As a result, the mass of the DM candidate is 
\be
m^2_{\va_{2}}= \fr 1 2 v_\eta^2
\la^{\eta\va}_2.   \label{eq9193}
\ee
According \cite{Tanabashi:2018oca}, the WIMP candidate has mass around 10 GeV, implying that $\la^{\eta\va}_2  \approx 0.04$.  To get the second DM candidate, namely, $\va^0_{1}$, we have to
carefully choose conditions.

Eqs. (\ref{eq752}) and (\ref{eq754}) result in a new complex $w$ defined as follows
\be
\om = \fr{1}{\sqrt{2}}(H_1 - i A_1) \, ,  \quad m^2_\om = \fr{\la_8 v^2_\chi}{2 \cos^2\theta_a } \, . \label{eqhay2}
\ee
\een
Let us rewrite the Higgs content in terms of the mass eigenstates mentioned above:
\begin{align}
\chi & \simeq
\begin{pmatrix}
G_{X^0} \\
G_{Y^-} \\
\fr{1}{\sqrt{2}}(v_\chi + R_{\chi_3^0} - i G_{Z^\prime })%
\end{pmatrix}
, \quad \rho =%
\begin{pmatrix}
\rho _{1}^{+} \\
\fr{1}{\sqrt{2}}(R_{\rho }- i I _{\rho }) \\
\rho _{3}^{+}%
\end{pmatrix}
,   \quad 
\eta   \simeq
\begin{pmatrix}
\fr{1}{\sqrt{2}}(v_\eta + h - i G_Z) \\
G_{W^-} \\
\om%
\end{pmatrix}
,  \crn
\va_2^{0}& = \fr{1}{\sqrt{2}} ( R_{\va_2} - i I_{\va_2})\sim
(1,1,0,i,1,0) \sim  \text{DM candidate}  ,\,  \crn
\xi^0& = \fr{1}{\sqrt{2}}(v_\xi + R_{\xi^0} - i G_M) \sim (1,1,0) \, .
\label{scalarsectorf}
\end{align}

\subsection{Simplified solutions}

We have shown that  the mass eigenstates of scalars  have been determined explicitly, except those relating to the two  $3\times3$ matrices \eqref{eq761} and \eqref{cpeven3}.  By introducing some further constrained  assumptions to simplify these matrices so that the physical states can be found, we will point out that  the parameter space of  the model under consideration contains valid regions, which are consistent with the experiment data including the $\rho$ parameter.  To reduce the arbitrary of  the unknown Higgs couplings in the potential  (\ref{ct2}), the following  relation are assumed firstly 
\be \la_{1}  = \la_2  \, , \hs
\la_{15} = \la_{16} \, , \hs
\la_{19} = \la_{20} \, ,\hs
w_1 = w_4 \, . \label{eqr1}
\ee 
In the next steps, we just pay attention to  find the masses and mass eigenstates of the two matrices \eqref{eq761} and \eqref{cpeven3}.  The other  will be summarized if necessary.

\subsubsection{The CP-odd  Higgs bosons}
Under  the assumption    (\ref{eqr1}),  the CP-odd scalar sector consists of four massless fields $\{I_{\chi_3^0}\,, I_{\eta}$, $G_M$, $G_1\}$  and four massive fields $\{A_1, A_2, A_3 I_{\va_2}\}$, as  summarized in Table \ref{masscpodd}.
\begin{table}[ht]
	\caption{Squared mass of CP-odd scalars under condition in (\ref{eqr1}) and $v_\chi \gg v_\eta$. }
	\resizebox{15cm}{!}{
		\renewcommand{\arraystretch}{1.2}
		\begin{tabular}{|c|c|c|c|c|c|c|c|c|}
			\hline
			Fields & $I_{\chi_1^0}=G_1\in G_{X^0}$ &$I_{\chi_3^0}=G_{Z'}$ & $I_{\eta_1^0}= G_Z$ &$I_{\eta_3^0}=A_1$
			& $I_\rho =A_2$ & $I_{\va_{1}^0}=A_3$ & $I_{\va_2 ^0} = DM$ & $I_{\xi^0}= G_M$
			\\ \hline
			Squared mass & $0$ & $0$ & $0$ & $m^2_{A_1}$ & $m^2_{A_2}$ & $m^2_{A_3}$ & $m^2_{I_{\va_2^0}}$ & $0$  \\
			\hline
	\end{tabular}}%
	\vspace{-0.1cm}
	\label{masscpodd}
\end{table}

\subsubsection{The  CP-even and SM-like Higgs bosons}

Now we turn to the sector where the SM Higgs boson exists, i.e.,  - the matrix in the basis $(R_{\chi_3^0}\, , R_{\eta_1^0}\, , R_{\xi^0})$ is given by Eq.~\eqref{cpeven3}.
Let us assume a simplified scenario worth to be considered is characterized
by the following relations:
\be
\la _{5}=\la _{13}=\la _{17}=\la _{\xi }=\la _{\chi \xi
}=\la _{\eta \xi }=\la ,\cm \cm v_{\xi }=v_{\chi }.
\label{simplifiedscenario}
\ee
In this scenario, the squared matrix (\ref{cpeven3}) takes the simple form:
\be \label{cpeven3a}
m_{CPeven3}^2 =\la \left(
\begin{array}{ccc}
2x^2  & x & x \\
x & 2 & 1 \\
x & 1 & 2%
\end{array}%
\right) v_{\chi }^2 \,,\cm \cm x=\fr{v_{\eta }}{v_{\chi }}=\tan\theta_a\, .
\ee
 Because  $v_{\chi } \gg v_{\eta }$, the  matrix \eqref{cpeven3a}  can
be perturbatively diagonalized as follows:
\be
R_{CPeven3}^{T}m_{CPeven3}^2 R_{CPeven3}\simeq \left(
\begin{array}{ccc}
\fr{4}{3}\la v_{\eta }^2  & 0 & 0 \\
0 & \la v_{\chi }^2  & 0 \\
0 & 0 & 3\la v_{\chi }^2 %
\end{array}%
\right), \quad R_{CPeven3}\simeq \left(
\begin{array}{ccc}
-1+\fr{x^2 }{9} & 0 & \fr{\sqrt{2}}{3}x \\
\fr{x}{3} & -\sqrt{\fr{1}{2}} & \sqrt{\fr{1}{2}} \\
\fr{x}{3} & \sqrt{\fr{1}{2}} & \sqrt{\fr{1}{2}}%
\end{array}%
\right) , \label{masshH45}
\ee
Thus, we find that the physical scalars included in the matrix $%
m_{CPeven3}^2 $ are:
\be
\left(
\begin{array}{c}
h \\
H_{4} \\
H_{5}%
\end{array}%
\right) \simeq \left(
\begin{array}{ccc}
-1+\fr{x^2 }{9} & \fr{x}{3} & \fr{x}{3} \\
0 & -\sqrt{\fr{1}{2}} & \sqrt{\fr{1}{2}} \\
\fr{\sqrt{2}}{3}x & \sqrt{\fr{1}{2}} & \sqrt{\fr{1}{2}}%
\end{array}%
\right) \left(
\begin{array}{c}
R_{\eta_1^0 } \\
R_{\chi_3^0 } \\
R_{\xi ^{0}}%
\end{array}%
\right) ,
\ee%
where $h$ is the SM-like Higgs boson with mass $126$ GeV identified with that found by LHC, whereas $H_{4}$ and $H_{5}$
are physical heavy scalars acquiring masses at the breaking scale of the $%
SU(3)_{L}\times U(1)_{X}\times Z_{4}\times Z_2 \times U(1)_{L_{g}}$
symmetry. Thus, we find that  $h$ has couplings very
close to SM expectation with small deviations of the order of $\fr{v_{\eta
}^2 }{v_{\chi }^2 }\sim\mathcal{O}(10^{-3})$.  In addition,  the squared masses of the physical scalars $h$ and $H_{4,5}$ are given in \eqref{masshH45}. 

Now, the content of the CP-even scalar sector is summarized in Table \ref{masscpeven}.
\begin{table}[ht]
\caption{Squared masses of CP-even scalars under condition in (\ref{simplifiedscenario}) and $v_\chi \gg v_\eta$. }
\resizebox{15cm}{!}{
\renewcommand{\arraystretch}{1.2}
\begin{tabular}{|c|c|c|c|c|c|c|c|c|}
\hline
Fields & $R_{\chi_1^0}\in G_{X^0}$ &$R_{\chi_3^0} \simeq H_4 $ & $R_{\eta_1^0} = h $ &$R_{\eta_3^0}=H_1$ & $R_\rho =H_2$ & $R_{\va_1^0}=H_3$ & $R_{\va_2 ^0}=DM$ & $R_{\xi^0} \simeq H_5$
\\ \hline
Squared mass & $0$ & $\la v_\chi ^2$ & $\fr 4 3 \la v_\eta ^2$ & $m^2_{H_1} = m^2_{A_1}$ & $m^2_{H_2}$ & $m^2_{H_3}$ & $m^2_{R_{\va_2^0}} = m^2_{I_{\va_2^0}}$ & $3\la v_\chi^2$  \\
\hline
\end{tabular}}%
\vspace{-0.1cm}
\label{masscpeven}
\end{table}

Taking into account  mass of the SM Higgs boson equal 126 GeV,  from Table  \ref{masscpeven} we obtain $\la \approx 0.187$, which can be used to calculate masses of the $H_{4,5}$ once $v_{\chi}$ is fixed. 

\subsubsection{The  charged Higgs bosons}

The charged scalar
sector contains two massless fields: $G_{W^+}$ and $G_{Y^+}$ which are
Goldstone bosons eaten by the longitudinal components of the $W^+$ and $Y^+$ gauge bosons, respectively.
The other massive fields are $\phi_1^+, \phi_2^+ $ and $\phi_4^+$ with
respective masses  given in \eqref{chiggsm}. 

In the basis ($\rho_1^+\, ,\,\rho_3^+\, ,\,\phi_3^+$), 
 the squared mass matrix is given in \eqref{eq761}. 
Let us make effort to simplify this  matrix. Note that $\mu_\chi^2, \mu_{\eta}^2$, and $\mu_{\xi}^2$ can be derived using relations \eqref{conditionn} and \eqref{simplifiedscenario}.  In addition, it is reasonable to assume

\be
\mu_\rho^2   =  -   \fr{v_\chi^2}{2} (\la_{18} + \la_{\rho\xi}) \approx \mu^2_\eta \,  , \quad  
\mu^2_{\phi_3^+}  =   - \fr{v_\chi^2}{2}( \la^{\chi\phi}_2  +\la^{\phi\xi}_2 )\,,
\label{eq9228}
\ee
we obtain the simple form of the squared mass matrix of the charged Higgs bosons,
\bea
M^2_{chargeds} &=
 \left(
\begin{matrix}
 \fr 1 2 v^2_\eta (\la_6 + \la_9)
 & 0 &
\fr{\la_3}{2}v_\eta v_\chi \\
0 &  \fr{1}{2}\!\left(v_\chi^2  \la_7 + \la_6 v^2_\eta \right)  & \fr{1}{\sqrt{2}} v_\chi w_2 \\
\fr{\la_3}{2}v_\eta v_\chi & \fr{1}{\sqrt{2}} v_\chi w_2 &
\fr{1}{2}v_\eta^2
\la^{\eta\phi}_2
\end{matrix}
\right)\, \label{eq92239}
\eea
The matrix  \eqref{eq92239}  predicts that there may exist two light  charged Higgs bosons $H^{+}_{1,2}$ with masses at the electroweak scale  and the mass of $H_3^+$ which is mainly composed of $\rho_3^+$ is around 3.5 TeV.
In addition, the Higgs boson $H_1^+$ almost does not carry lepton number, whereas the others two do.

	Generally, the Higgs potential always contains mass terms which mix VEVs. However, these terms must be small enough to avoid  high order divergences (for examples, see Refs. \cite{Phong:2018kgv,Baker:2017zwx}) and provide baryon asymmetry of Universe by the strong electroweak phase transition (EWPT).
	
	Ignoring mixing term containing $\la_3$ in (\ref{eq92239}) does not  affect other physical aspects, since the above mentioned term just increases
	or decreases small amount of the charged Higgs bosons. Therefore, without lose of generality, neglecting the term with $\la_3$ satisfies other aims such as EWPT.
	
	Hence, in the matrix of (\ref{eq92239}), the coefficient  $\la_3$ is reasonably assumed to be zero. Therefore we get immediately   one physical field $\rho_1^+$ with mass given by
	\be m^2_{\rho_1^+} = \fr 1 2 v^2_\eta (\la_6 + \la_9) \, .
	\label{l101} 
	\ee
	The other fields mix by submatrix given at the bottom of (\ref{eq92239}).
 The limit $\rho_1^+=H^+_1$ when $\lambda_3=0$ is very interesting for discussion of the Higgs contribution to the $\rho$ parameter.

The content of the charged scalar sector is summarized in Table \ref{masscharged}.
\begin{table}[tbp]
	\caption{Squared mass of charged  scalars under condition in (\ref{eq9228}) and $v_\chi \gg v_\eta$. }
	\resizebox{15cm}{!}{
		\renewcommand{\arraystretch}{1.2}
		\begin{tabular}{|c|c|c|c|c|c|c|c|c|}
			\hline
			Fields & $\eta_2^+=G_{W^+}$ &$\chi_2^+ = G_{Y^+} $ & $H_1^+$ &$H_3^+$ & $H_2^+$ & $\phi_1^+$ & $\phi_2^+$ & $\phi_4^+$
			\\ \hline
			Squared mass & $0$ & $0$ & $m^2_{H_1^+}$ & $m^2_{H_3^+} $ & $m^2_{H_2^+}$ & $m^2_{\phi_1^+}$
			& $m^2_{\phi_2^+} $ & $m^2_{\phi_4^+}$  \\
			\hline
	\end{tabular}}%
	\vspace{-0.1cm}
	\label{masscharged}
\end{table}
It is worth mentioning that the masses of three charged scalars $\phi_i^+, i=1, 2, 4$ are still not fixed.

The potential including  lepton number violations,i.e., $V_{full} = V_{LNC} + V_{LNV}$  is quite similar to the
previous one. There are some differences:

\ben
	\item The masses of the fields receive some new contributions.
	
	\item The complex scalar $\va_2^0$ has the same mass in both cases.
	
	\item Majoron does not exist and its mass only arises from lepton number
	violating scalar interactions.
	
	\item The mixing of the CP-even scalar fields is more complicated.
\een
\subsection{Scalar contributions  to the $\rho$ parameter}
\begin{figure}[ht]
		\centering
		\begin{tabular}{cc}
			\includegraphics[width=7cm]{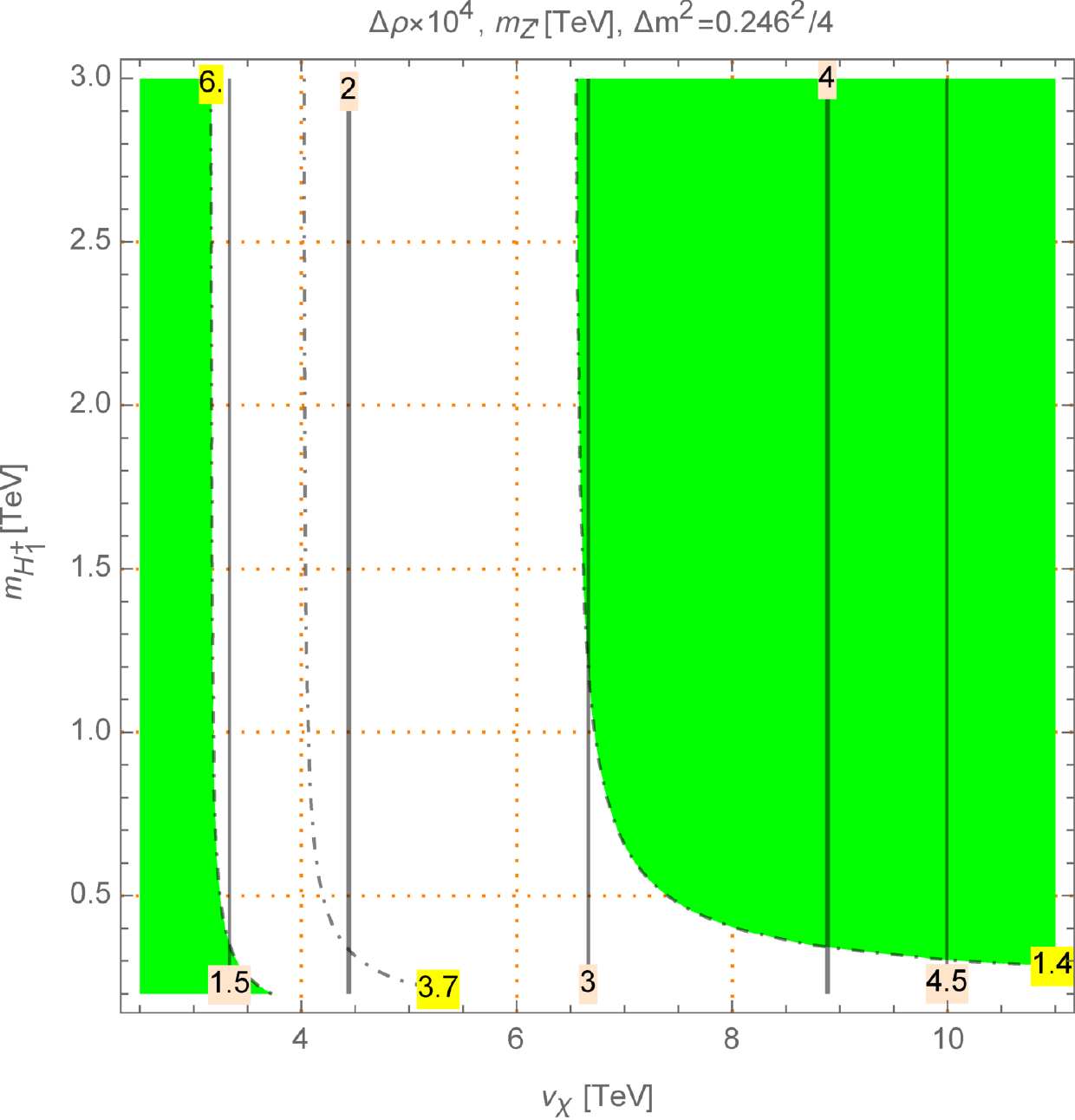} & %
			\includegraphics[width=7cm]{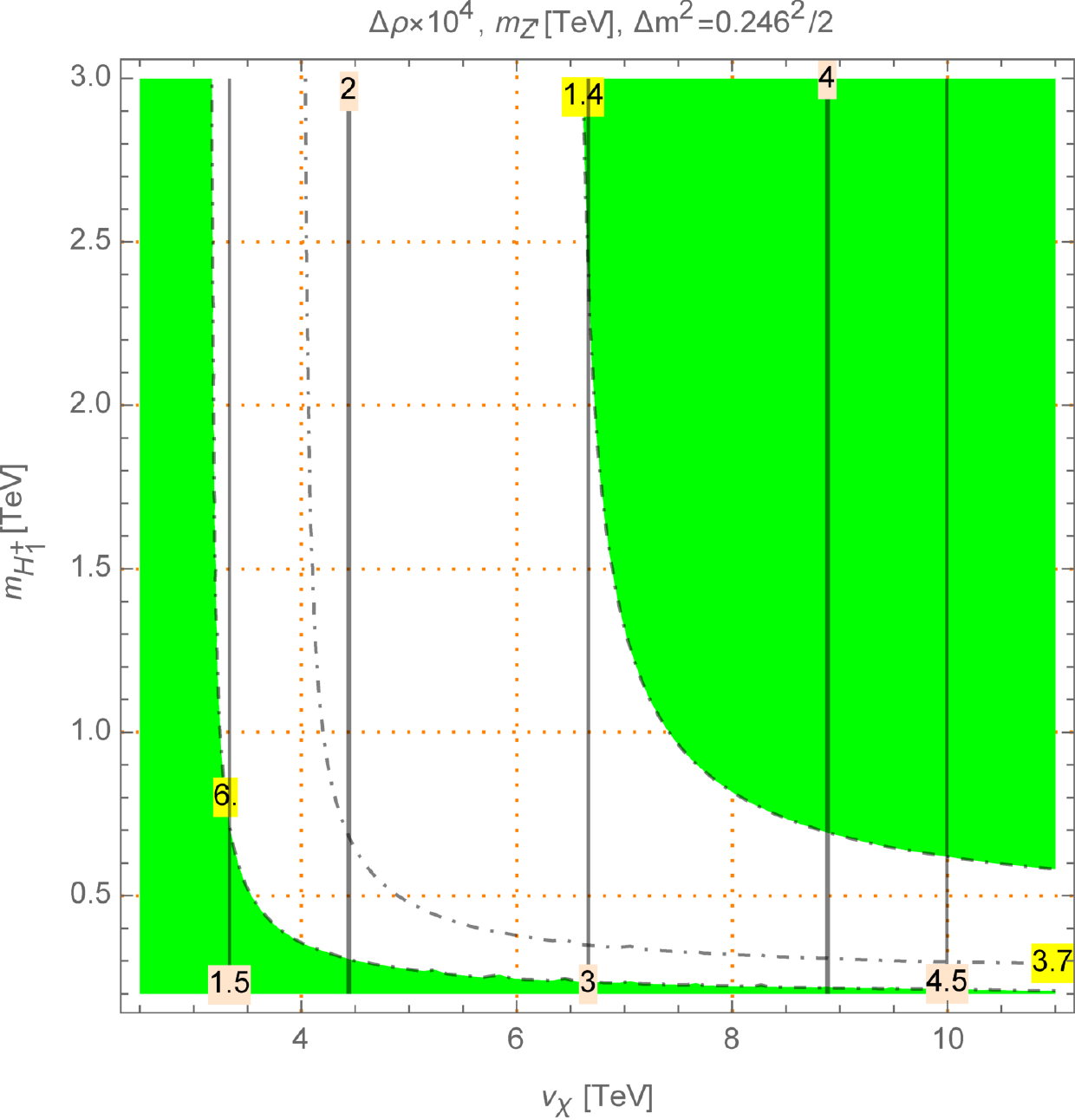} \\
			\includegraphics[width=7cm]{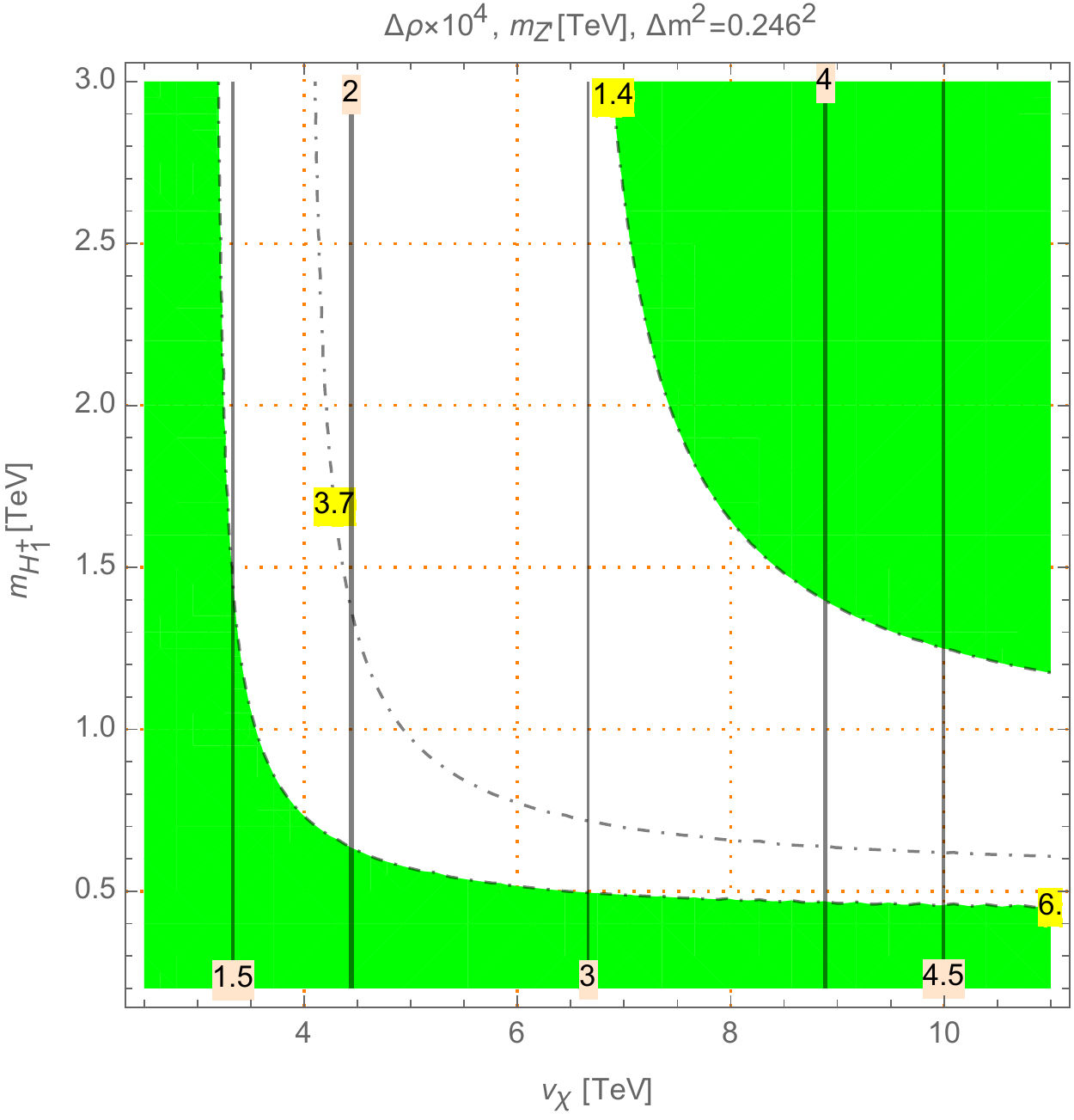} & %
			\includegraphics[width=7cm]{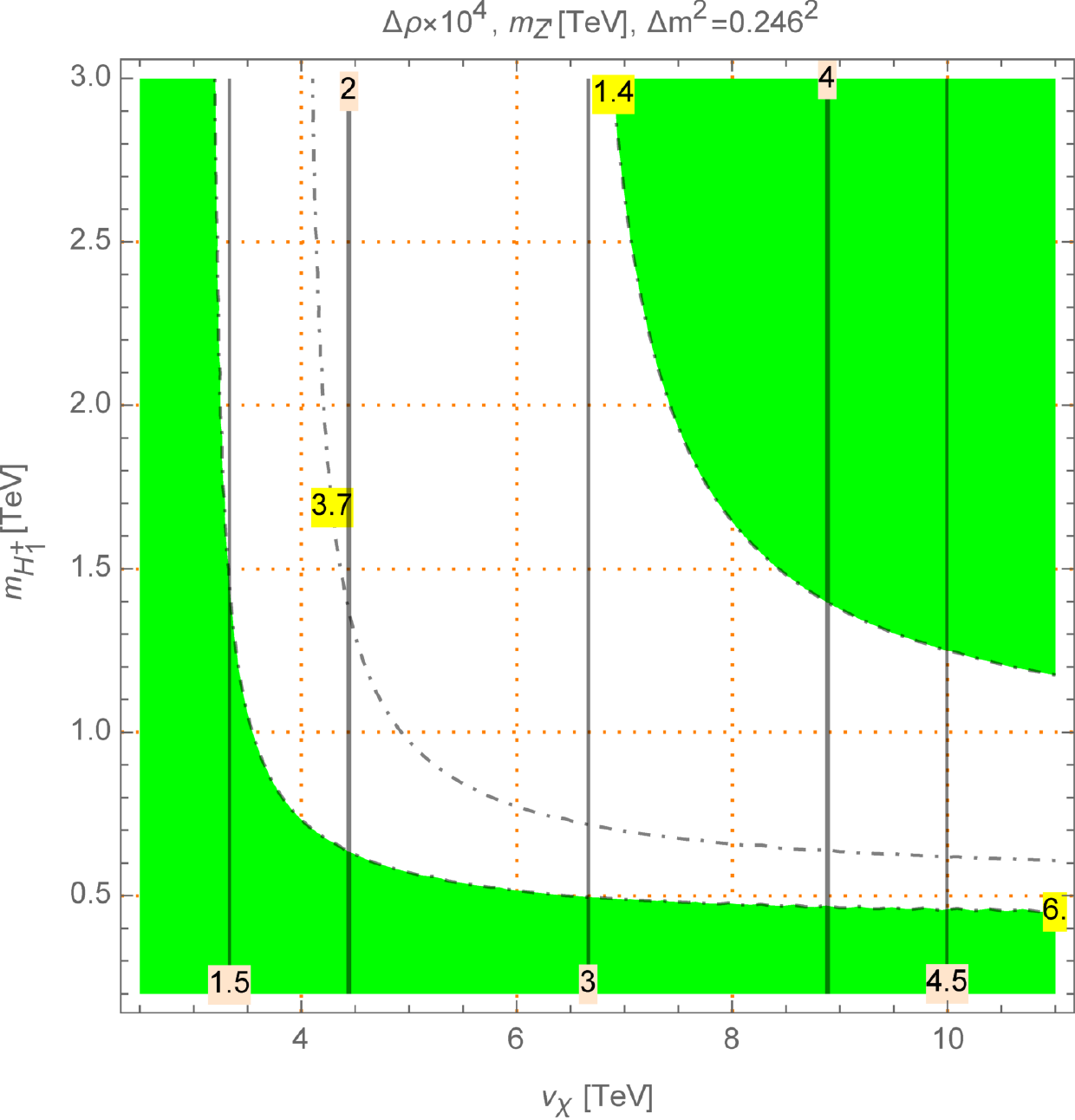} \\
			&
		\end{tabular}%
		\caption{Contour plots of the $\rho$ parameter (dotted-dashed curves) and $M_{Z'}$ (black currves) as  functions of $v_\chi$ and $m_{H^+_1}$. The green regions are excluded by the recent experimental constrain of the $\rho$.}
		\label{fig1ig_contour1}
	\end{figure}

	The new Higgs bosons may give contribution to the $\rho$ parameter at one-loop level, as shown in many models beyond the SM, such as the simplified models~\cite{Lopez-Val:2014jva}, the Two Higgs Doublet Models \cite{LopezVal:2012zb, Hernandez:2015dga}, and the supersymmetric version of the SM \cite{Drees:1990dx}. In the 3-3-1 CKS model, we will consider the effect of the Higgs contributions to the $\rho$ parameter at one-loop level.  These contributions will be determined in  the limit of the  suppressed  $Z-Z'$ mixing and the decoupling of the SM-like Higgs boson with other CP-even neutral Higgs bosons. As a consequence, the one-loop contribution of the SM-like Higgs boson to the $\rho$ parameter is the same as in the SM. Excepting for the components of the scalar triplet $\rho$, the other heavy CP-even neutral Higgs bosons do not couple with the SM gauge bosons $W$ and $Z$  and thus they do not provide contributions 
	to the $\rho$ parameter. Contributions of the remaining Higgs bosons can be calculated using the results given in Ref.~\cite{Drees:1990dx}. In particular, contributions of any Higgs bosons in our case to $\Delta\rho$ are determined as follows   
	\begin{equation}\label{eq_Deltarho }
	\Delta\rho=\frac{\Pi_{WW}(0)}{M_W^2} - \frac{\Pi_{ZZ}(0)}{M^2_Z},
	\end{equation}
	where $\Pi_{WW}(0)$ and $\Pi_{ZZ}(0)$ are the coefficients of $-i g_{\mu\nu}$ in the vacuum-polarization amplitudes of charged and neutral $W$ bosons and $Z$ gauge bosons, respectively. Our case relates with only the contribution of "non-Higgs scalars" $\phi_{1,2}$ with masses $m_{1,2}$  and coupling 
	\begin{equation}\label{eq_NonHiggs}
	ic\phi^*_1 \overset{\leftrightarrow}{\partial}_\mu \phi_2 V^{\mu} \equiv~ i c \left[\phi^*_1\partial_\mu \phi_2 - (\partial_\mu\phi^*_1) \phi_2 \right] V^{\mu}, \quad (V=W,Z), 
	\end{equation}
	The corresponding contribution is 
	\begin{equation}\label{eq_piscalar}
	\Pi(\mathrm{scalar }) =\frac{|c|^2}{16\pi^2} f_s(m_1,m_2),
	\end{equation} 
	where 
	\begin{align}
	\label{eq_fsm12}
	f_s(m_1,m_2)&= f_s(m_2,m_1)= \frac{m_1^2m_2^2}{m_1^2-m_2^2}\ln\left[ \frac{m_2^2}{m_1^2} \right] +\frac{1}{2} \left(m_1^2 +m_2^2\right)\crn
	&= m_1^2f_s(x)=m_1^2 \left( \frac{x\ln(x)}{1-x} +\frac{1+x}{2}\right), \quad x_{21}\equiv \frac{m_2^2}{m_1^2}. 
	\end{align}
	The function in Eq.~\eqref{eq_fsm12} satisfies  $f_s(m_1,m_1)=\underset{m_2\rightarrow m_1}{\lim} f_{s}(m_1,m_2)=0$ and $f_s(m_1,m_2)>0$ with $m_1\neq m_2$.  As a consequence, the charged Higgs bosons $\phi^{\pm}_{1,2,4}$ having vanishing Higgs-gauge couplings with other Higgs bosons give vanishing contributions to the $\rho$ parameter. Nonvanishing contributions now may arise from the charged \Long{Higgs bosons} $H^{\pm}_{1,2,3}$ corresponding to the basis ($\rho^{\pm}_1,\rho^{\pm}_3,\phi^{\pm}_3$)
	and the CP-odd neutral Higgs relating with $I_{\rho}$. The relevant Lagrangian is 
	\begin{align} \label{eq_LVHH}
	\mathcal{L}_{VHH}&=(D_{\mu}\rho)^{\dagger}(D^{\mu}\rho) + (D_{\mu}\phi^+_3)^{*}(D^{\mu}\phi^+_3) \crn  
	\rightarrow& \frac{ig}{2} Z^{\mu} \left[ \frac{1-2s^2_W}{c_W} \left( \rho^-_1 \overset{\leftrightarrow}{\partial}_\mu\rho^+_1 \right) + \frac{-2s^2_W}{c_W} \left( \rho^-_3 \overset{\leftrightarrow}{\partial}_\mu\rho^+_3 \right)  + \frac{-i}{c_W} \left( I_{\rho} \overset{\leftrightarrow}{\partial}_\mu R_{\rho}\right) \right] \crn 
	& + \frac{-igs^2_W}{c_W} Z^{\mu} \left( \phi^-_3 \overset{\leftrightarrow}{\partial}_\mu\phi^+_3 \right) +\left[   -\frac{ig}{2} W^{+\mu}  \rho^-_1 \left( \overset{\leftrightarrow}{\partial}_\mu R_{\rho} - i  \overset{\leftrightarrow}{\partial}_\mu I_{\rho} \right) +\mathrm{H.c.}\right]. 
	\end{align} 
	In the scalar-gauge interactions of Eq~\eqref{eq_LVHH}, only the last term contributes to $\Pi_{WW}(0)$, thus giving rise to a 	 non-negative contribution to the $\rho$ parameter,  which may make the allowed $M_{Z'}$ mass to move  outside the excluded region recently reported by  LHC searches~\cite{Aaboud:2017sjh}.  On the contrary, all of the remaining terms contributing 
	 to $\Pi_{ZZ}(0)$, give non-positive contributions to the $\rho$ parameter.  For illustration, we will consider a simple case where only positive contributions to the $\rho$ parameter are kept, namely $\rho^{\pm}_{1,3}\equiv H^{\pm}_{1,3}$,  $\phi^{\pm}_3\equiv H^{\pm}_2$, $I_{\rho}\equiv A_{2}$ and $R_{\rho}$ are 
	 mass eigenstates. Then, all contributions to $\Pi_{ZZ}(0)$ arising from the charged Higgs bosons are proportional to $f_s(m_s,m_s)=0$ with $s=\rho^{\pm}_{1,3}, \phi^{\pm}_3$. In addition, the simplified condition \eqref{eqr1} with $\lambda_1 \ll 1$ results in $m^2_{I_{\rho}}=m^2_{R_{\rho}}$, leading to a vanishing neutral Higgs boson contribution to $\Pi_{ZZ}(0)$:  $f_s(m_{I_{\rho}},m_{R_{\rho}})=0$. The only non-zero contribution has the form  
	\begin{equation}\label{droRIrho}
	\Delta\rho^{H}= \frac{g^2}{16\pi^2 m^2_W} f_{s}(m_{H^+_1},m_{R_\rho})= \frac{\sqrt{2}G_F}{16\pi^2 } f_{s}(m_{H^{+}_1},m_{R_\rho}),
	\end{equation}
	where 
	\begin{equation}\label{eq_dm2RhoHp1}
	\Delta m^2\equiv m^2_{R_{\rho}}-m_{H^+_1}= -\frac{\lambda_9 v^2_{\eta}}{2}\sim \mathcal{O}(v^2_{\eta}). 
	\end{equation}
	Allowed regions of the parameter space for some specific values of $\Delta m^2$ are shown in Fig.~\ref{fig1ig_contour1}. 
	It can be seen that the  large values of $v_{\chi}$ are still allowed, thus implying that
	no upper bounds are required.  The allowed values of $v_{\chi}$ and $M_{Z'}$ strongly depend on the lower bound of $m_{H^{\pm}_1}$ and $m_{R_{\rho}}$, which may have previously reported from LHC searches. Unfortunately,  the Higgs triplets $\rho$ containing the neutral components $\rho^0_2$ with zero VEV, hence all of the three components of $\rho$  do not couple to the two SM gauge bosons $Z$ and $W$. This Higgs triplet also does not contribute to the SM-like Higgs boson. As a result, all of the  Higgs bosons $H^{\pm}_1$, $R_{\rho}$, and $I_{\rho}$ are not affected by  the following decays searched by LHC:  $H^{\pm}_1\rightarrow W^{\pm} Z,W^{\pm} h$ and $R_{\rho}, I_{\rho}\rightarrow W^+W^-, ZZ, Zh$.  These Higgs bosons do not couple with the SM quarks \cite{CarcamoHernandez:2017cwi},   can not be produced at LHC from the recent chanel searching ~\cite{Aaboud:2018gjj}.  Only the  allowed tree level decays to two SM fermions  are leptonic decays:  $H^{\pm}_1\rightarrow \nu_{2,3} \tau, \nu_{2,3} \mu $  and $R_{\rho}, I_{\rho}\rightarrow \bar{e}_i e_i$ ($i=1,2,3$), which are also searched by LHC, but the couplings of these Higgs with SM quarks are necessary to produce these Higgs bosons .  
	
	Because the above Higgs bosons have couplings with many new charged particles in the model under consideration, their one-loop level decays to photons may appear, thus making them interesting  channels for their search at the LHC, in particular these charged Higgs bosons feature the following decay modes $H^{\pm}_1\rightarrow W^{\pm} \gamma$,  $R_{\rho}, I_{\rho}\rightarrow Z\gamma$~\cite{Aaboud:2018fgi}, and $R_{\rho}, I_{\rho}\rightarrow \gamma\gamma$~\cite{Khachatryan:2016yec,Aaboud:2017yyg}. The heavy neutral Higgs boson masses are predicted to be at the TeV scale, which is outside the LHC excluded regions.  Combined with the relation \eqref{eq_dm2RhoHp1}, the mass of the charged Higgs boson $H^+_1$ should also be at the TeV scale. From the figure~\ref{fig1ig_contour1}, we can see that $M_{Z'}\ge 4 $ TeV is allowed if $\Delta m^2 $ is large enough, for example $\Delta m^2\ge (0.246 \,\mathrm{TeV})^2$.

\section{Charged lepton flavor violating decay constraints.}
\label{LFV} 
\begin{figure}[tbp]
\includegraphics[width=0.85\textwidth]{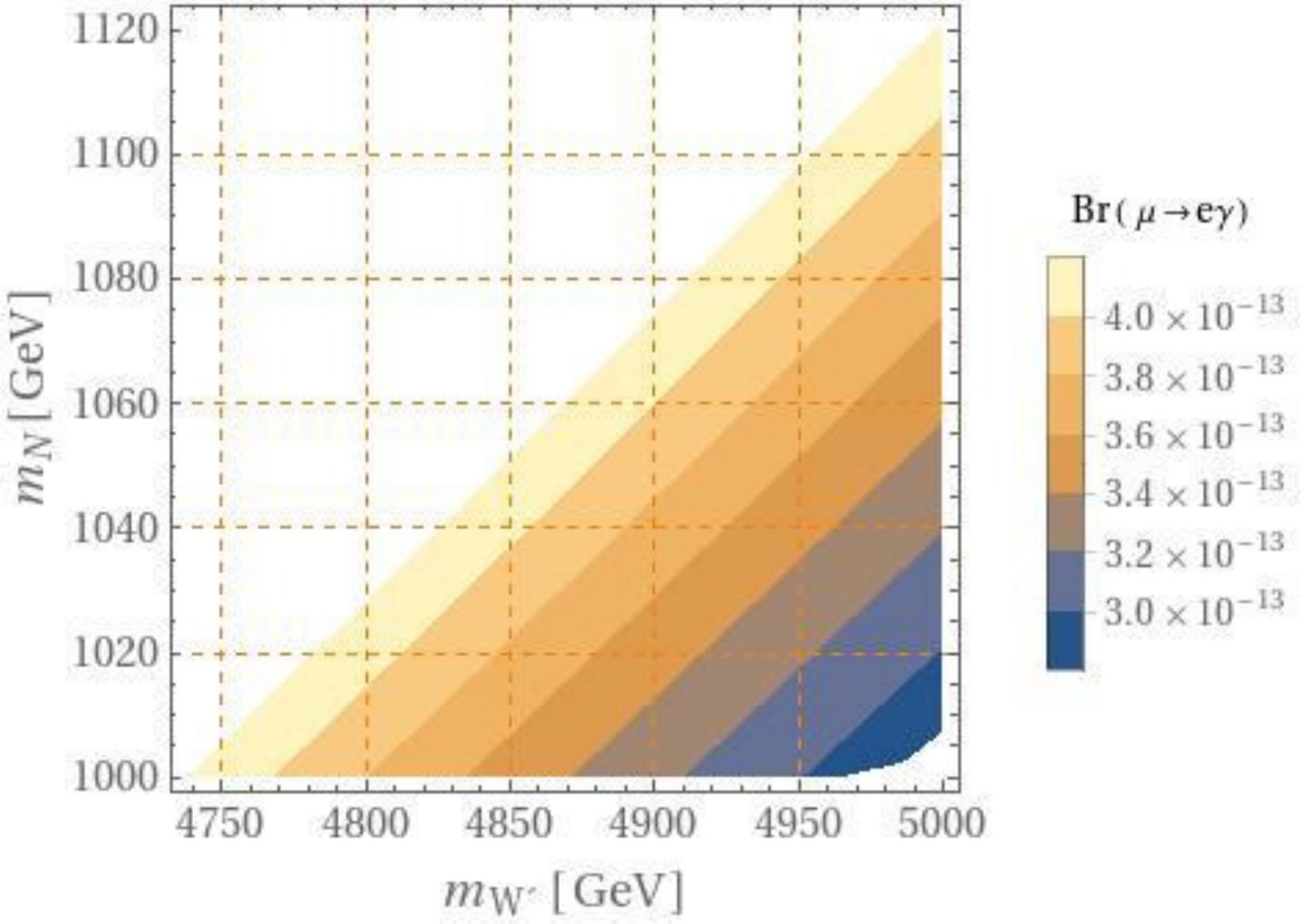}
\caption{Allowed parameter space in the $m_{W^{\prime }}-m_{N}$ plane consistent with the LFV constraints.}
\label{LFVplot}
\end{figure}

In this section we will determine the constraints that the
charged lepton flavor violating decays $\mu \rightarrow e\gamma$, $\tau
\rightarrow \mu \gamma $ and $\tau \rightarrow e\gamma $ imposed on the
parameter space of our model. As mentioned in Ref. \cite{CarcamoHernandez:2017cwi}, the sterile neutrino
spectrum of the model is composed of two almost degenerate neutrinos with
masses at the Fermi scale and four nearly degenerate neutrinos with TeV scale
masses. These sterile neutrinos together with the heavy $W^{\prime }$ gauge
boson induce the $l_{i}\rightarrow l_{j}\gamma $ decay at one loop level,
whose branching ratio is given by: \cite{Ilakovac:1994kj,Deppisch:2004fa,Lindner:2016bgg}: 
\begin{eqnarray}
Br\left( l_{i}\rightarrow l_{j}\gamma \right)  &=&\frac{\alpha
_{W}^{3}s_{W}^{2}m_{l_{i}}^{5}}{256\pi ^{2}m_{W^{\prime }}^{4}\Gamma _{i}}%
\left\vert 2G\left( \frac{m_{N_{1}}^{2}}{m_{W^{\prime }}^{2}}\right)
+4G\left( \frac{m_{N_{2}}^{2}}{m_{W^{\prime }}^{2}}\right) \right\vert ^{2},%
\hspace{0.5cm}\hspace{0.5cm}\hspace{0.5cm},  \notag \\
G\left( x\right)  &=&-\frac{2x^{3}+5x^{2}-x}{4\left( 1-x\right) ^{2}}-\frac{%
3x^{3}}{2\left( 1-x\right) ^{4}}\ln x.  \label{Brmutoegamma}
\end{eqnarray}%
In our numerical analysis we have fixed $m_{N_{1}}=100$ GeV and we have
varied the $W^{\prime }$ gauge boson mass in the range $4$ TeV$\lesssim
m_{W^{\prime }}\lesssim 5$ TeV. We consider neutral heavy $Z^\prime $ gauge boson masses larger than $4$ TeV to fullfill the bound arising from the experimental data on $K$, $D$ and $B$ meson mixings \cite{Huyen:2012uk}. Figure \ref{LFVplot} shows the allowed parameter
space in the $m_{W^{\prime }}-m_{N}$ plane consistent with the constraints
arising from charged lepton flavor violating decays. As seen from Figure \ref%
{LFVplot}, the obtained values for the branching ratio of $\mu \rightarrow
e\gamma $ decay are below its experimental upper limit of $4.2\times 10^{-13}
$ since these values are located in the range $3\times 10^{-13}\lesssim
Br\left( \mu \rightarrow e\gamma \right) \lesssim 4\times 10^{-13}$, for 
sterile neutrino masses $m_{N_{2}}$ lower than about $1.12$ TeV. In the same
region of parameter space, the obtained branching ratios for the $\tau
\rightarrow \mu \gamma $ and $\tau \rightarrow e\gamma $ decays can reach
values of the order of $10^{-13}$, which is below their corresponding upper
experimental bounds of $4.4\times 10^{-9}$ and $3.3\times 10^{-9}$,
respectively. Consequently, our model is compatible with the charged lepton
flavor violating decay constraints provided that the sterile neutrinos are
lighter than about $1.12$ TeV. 

\section{Search for $Z^\prime$ at LHC}
\label{collider}

In this section, we present two typical effects of the LHC, namely, production of single a particle in proton-proton collisions.

\subsection{Phenomenology of $Z^\prime$ gauge boson}
\label{zlhc}

In what follows we proceed to compute the total cross section
for the production of a heavy $Z^\prime $ gauge boson at the LHC via {%
 Drell-Yan mechanism.  In our computation for the total cross
section we consider the dominant contribution due to the parton distribution
functions of the light up, down and strange quarks, so that the total cross
section for the production of a $Z^\prime $ via quark antiquark
annihilation in proton proton collisions with center of mass energy $\sqrt{S}
$ takes the form:
\small{
\bea
\si _{pp\rightarrow Z^\prime }^{\left( DrellYan\right) }( S)
&=&\fr{g^2 \pi }{6c_{W}^2 S}\left\{ \left[ \left( g_{uL}^{\prime
}\right) ^2 +\left( g_{uR}^\prime \right) ^2 \right] \int_{\ln \sqrt{%
\fr{m_{Z^\prime }^2 }{S}}}^{-\ln \sqrt{\fr{m_{Z^\prime }^2 }{S}}%
}f_{p/u}\left( \sqrt{\fr{m_{Z^\prime }^2 }{S}}e^{y},\mu ^2 \right)
f_{p/\overline{u}}\left( \sqrt{\fr{m_{Z^\prime }^2 }{S}}e^{-y},\mu
^2 \right) dy\right.  \crn
&&+\left. \left[ \left( g_{dL}^\prime \right) ^2 +\left( g_{dR}^{\prime
}\right) ^2 \right] \int_{\ln \sqrt{\fr{m_{Z^\prime }^2 }{S}}}^{-\ln
\sqrt{\fr{m_{Z^\prime }^2 }{S}}}f_{p/d}\left( \sqrt{\fr{m_{Z^{\prime
}}^2 }{S}}e^{y},\mu ^2 \right) f_{p/\overline{d}}\left( \sqrt{\fr{%
m_{Z^\prime }^2 }{S}}e^{-y},\mu ^2 \right) dy\right.  \crn
&&+\left. \left[ \left( g_{dL}^\prime \right) ^2 +\left( g_{dR}^{\prime
}\right) ^2 \right] \int_{\ln \sqrt{\fr{m_{Z^\prime }^2 }{S}}}^{-\ln
\sqrt{\fr{m_{Z^\prime }^2 }{S}}}f_{p/s}\left( \sqrt{\fr{m_{Z^{\prime
}}^2 }{S}}e^{y},\mu ^2 \right) f_{p/\overline{s}}\left( \sqrt{\fr{%
m_{Z^\prime }^2 }{S}}e^{-y},\mu ^2 \right) dy\right\}
\nn
\eea%
}

\begin{figure}[tbh]
\resizebox{12cm}{10cm}{\includegraphics{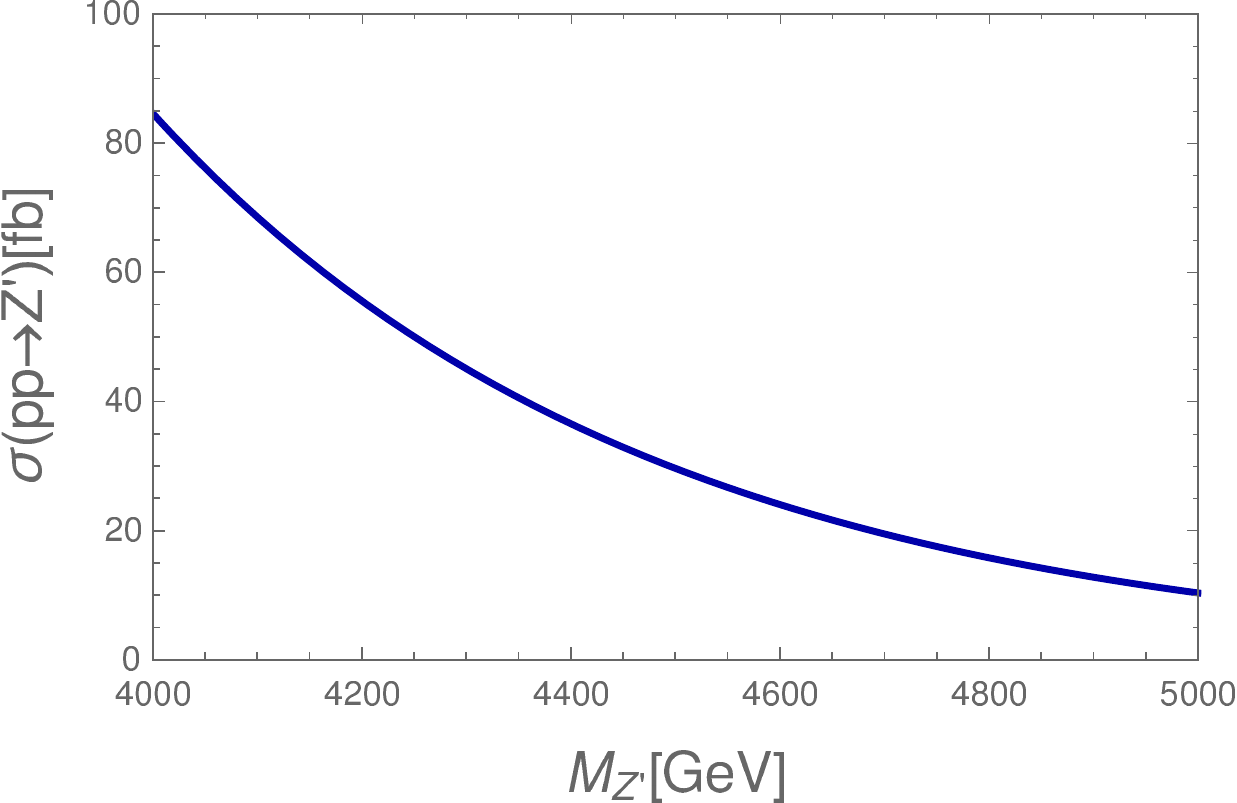}}\vspace{0cm}
\caption{Total cross section for the $Z^\prime $ production via Drell-Yan
mechanism at the LHC for $\protect\sqrt{S}=13$ TeV and as a function of the $%
Z^\prime $ mass.}
\label{qqtoZprime}
\end{figure}
Figure \ref{qqtoZprime} displays the $Z^\prime $ total production cross section at the
LHC via Drell-Yan mechanism at the LHC for $\protect\sqrt{S}=13$ TeV and as a function of the $%
Z^\prime $ mass, which is taken to range from $4$ TeV up to $5$ TeV. We consider neutral heavy $Z^\prime $ gauge boson masses larger than $4$ TeV to fullfill the bound arising from the experimental data on $K$, $D$ and $B$ meson mixings \cite{Huyen:2012uk}. For such as a region of $Z^\prime $ masses we find that the total production cross section ranges from $85$ fb up to $10$ fb. 
The heavy neutral $Z^\prime$ gauge boson after being produced it will decay into pair of SM particles, with dominant decay mode into quark-antiquark pairs as shown in detail in Refs. \cite{Perez:2004jc,CarcamoHernandez:2005ka}. The two body decays of the $Z^\prime$ gauge boson in 3-3-1 models have been studied in details in Refs. \cite{Perez:2004jc}. In particular, in Ref. \cite{Perez:2004jc} it has been shown the $Z^\prime$ decays into a lepton pair in 3-3-1 models have branching ratios of the order of $10^{-2}$, which implies that the total LHC cross section for the $pp\to Z^\prime\to l^{+}l^{-}$ resonant production at $\protect\sqrt{S}=13$ TeV will be of the order of $1$ fb for a $4$ TeV $Z^\prime$ gauge boson, which is below its corresponding lower experimental limit arising from LHC searches \cite{Aaboud:2017sjh}. On the other hand, at the proposed energy upgrade of the LHC at 28 TeV center of mass
energy, 
the total cross section for the Drell-Yan production of
  a heavy $Z^\prime $ neutral gauge boson gets significantly enhanced reaching values ranging from $2.5$ pb up to $0.7$ pb,
as indicated in figure \ref{qqtoZprimefor28TeV}. Consequently, the LHC cross section for the $pp\to Z^\prime\to l^{+}l^{-}$ resonant production at $\protect\sqrt{S}=28$ TeV will be of the order of $10^{-2}$ pb for a $4$ TeV $Z^\prime$ gauge boson, which corresponds to the order of magnitude of its corresponding lower experimental limit arising from LHC searches \cite{Aaboud:2017sjh}.
\begin{figure}[tbh]
\resizebox{12cm}{10cm} {\includegraphics{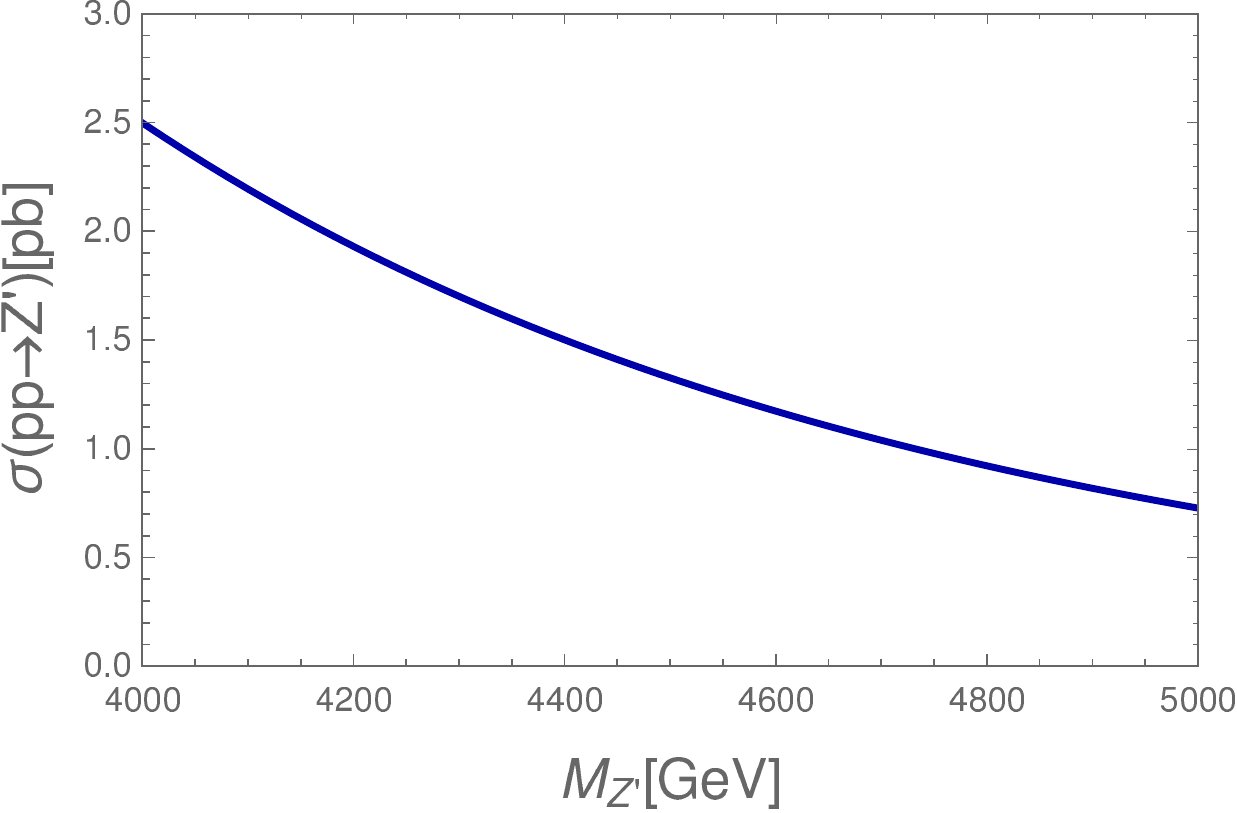}}\vspace{0cm}
\caption{Total cross section for the $Z^\prime $ production via Drell-Yan mechanism at the proposed energy upgrade of the LHC with $\protect\sqrt{S}=28$ TeV as a function of the $Z^\prime $ mass.}
\label{qqtoZprimefor28TeV}
\end{figure}

\subsection{Phenomenology of $H_{4}$ Heavy Higgs boson}
\label{h4lhc}

In what follows we proceed to compute the LHC production cross section of
the singly heavy scalar $H_{4}$. Let us note that the singly heavy scalar $%
H_{4}$ is mainly produced via gluon fusion mechanism mediated by a
triangular loop of the heavy exotic quarks $T$, $J_1$ and $J_2$. Thus,
the total cross section for the production of the heavy scalar $H_{4}$
through gluon fusion mechanism in proton proton collisions with center of
mass energy $\sqrt{S}$ takes the form:
\bea
\si _{pp\rightarrow gg\rightarrow H_{4}}( S) &=&\fr{\al
_{S}^2 m_{H_{4}}^2 \left\vert \left( R_{CPeven3}\right) _{22}\right\vert
^2 }{64\pi v_{\chi }^2 S}\left[ I\left( \fr{m_{H_{4}}^2 }{m_{T}^2 }%
\right) +I\left( \fr{m_{H_{4}}^2 }{m_{J_{1}}^2 }\right) +I\left( \fr{%
m_{H_{4}}^2 }{m_{J_2 }^2 }\right) \right]  \crn
&&\times \int_{\ln \sqrt{\fr{m_{H_{4}}^2 }{S}}}^{-\ln \sqrt{\fr{%
m_{H_{4}}^2 }{S}}}f_{p/g}\left( \sqrt{\fr{m_{H_{4}}^2 }{S}}e^{y},\mu
^2 \right) f_{p/g}\left( \sqrt{\fr{m_{H_{4}}^2 }{S}}e^{-y},\mu
^2 \right) dy
\nn
\eea
where $f_{p/g}\left( x_1,\mu ^2 \right) $ and $f_{p/g}\left(x_2,\mu
^2 \right) $ are the distributions of gluons in the proton which carry
momentum fractions $x_1$ and $x_2$ of the proton, respectively. Furthermore $%
\mu =m_{H_{4}}$ is the factorization scale and $I( z) $ is given
by:
\be
I( z) =\int_{0}^{1}dx\int_{0}^{1-x}dy\fr{1-4xy}{1-zxy}
\label{g1a}
\ee
\begin{figure}[tbh]
\resizebox{12cm}{10cm}{\includegraphics{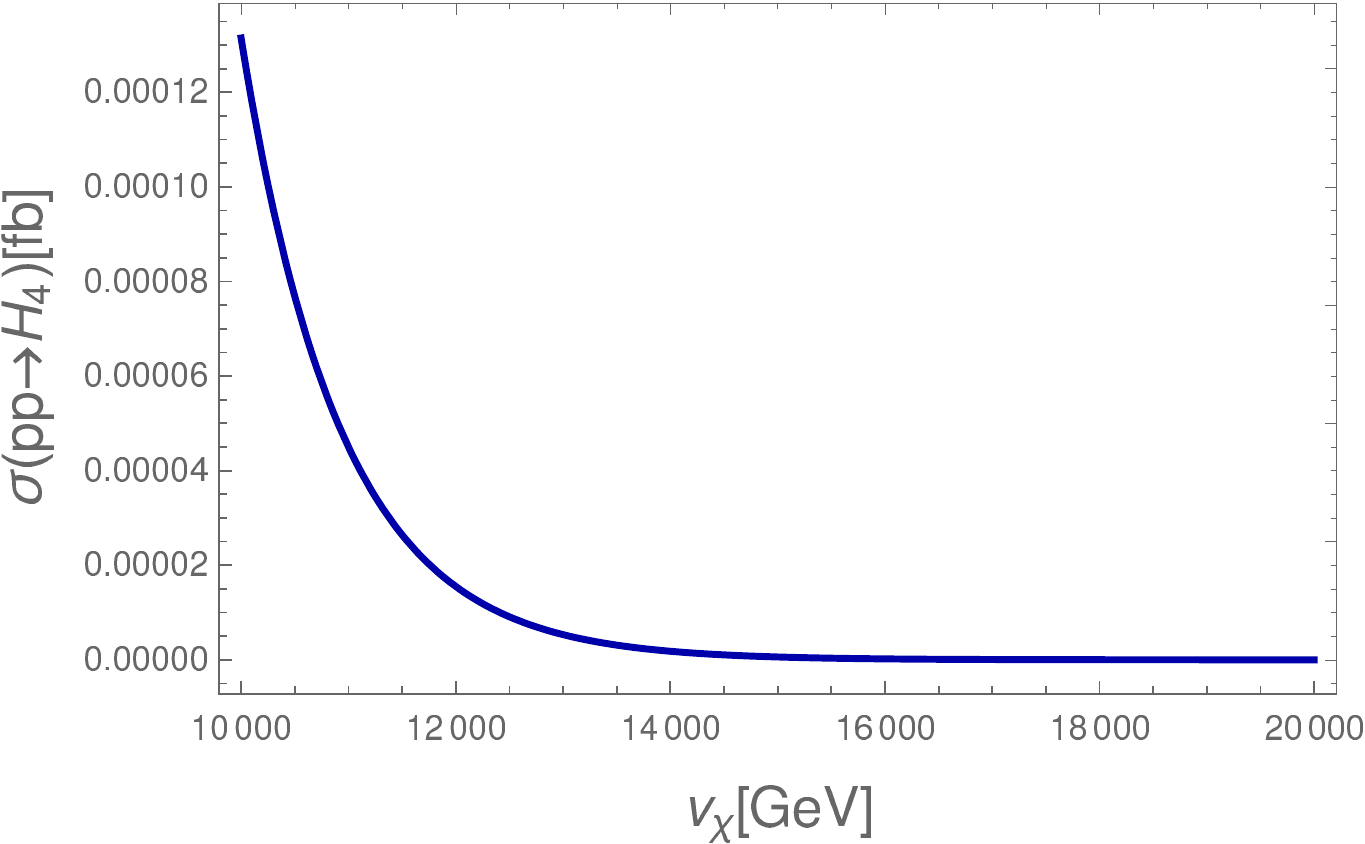}}\vspace{0cm}
\caption{Total cross section for the $H_4$ production via gluon fusion
mechanism at the LHC for $\protect\sqrt{S}=13$ TeV and as a function of the $%
SU(3)_L\times U(1)_X$ symmetry breaking scale $v_{\protect\chi}$ for the
simplified scenario described in Eq. (\ref{simplifiedscenario})}.
\label{ggtoH4}
\end{figure}
Figure \ref{ggtoH4} displays the $H_4$ total production cross section at the
LHC via gluon fusion mechanism for $\sqrt{S}=13$ TeV, as a function of the $%
SU(3)_L\times U(1)_X$ symmetry breaking scale $v_\chi$, which is taken to
range from $10$ TeV up to $20$ TeV. The aforementioned range of values
for the $SU(3)_L\times U(1)_X$ symmetry breaking scale $v_\chi$
corresponds to a heavy scalar mass $m_{H_4}$  varying between $4.4$ TeV and $8.9$ TeV. Considering the mass of the heavy scalar field $H_4$ in the range $8$ TeV $\lesssim$ $M_{H_4}$$\lesssim$ $8.9$ TeV,  
	it is reasonable to assume that it will have dominant decay modes into $W^{\prime}W^{\prime}$ and $Z^{\prime}Z^{\prime}$ heavy gauge boson pairs. On the other hand, for a heavy scalar field $H_4$ with mass in the range $4.4$ TeV $\lesssim$ $M_{H_4}$$\lesssim$ $8$ TeV, based on Ref. \cite{Branco:2011iw}, it is reasonable to assume that its dominant decay mode will be on $t\bar{t}$ pair. Furthermore, in the region of $H_4$ masses considered in our analysis, the $H_4$ decay into exotic quark pairs will be kinematically forbidden for exotic quark Yukawa couplings of order unity.  Note that we have chosen values for $v_{\chi}$ larger than $10$ TeV, which corresponds to a $Z^{\prime}$ gauge boson heavier than $4$ TeV, which is required to guarantee the consistency of 331 models with the experimental data on $K$, $D$ and $B$ meson mixings \cite{Huyen:2012uk}.    Here, for the sake of simplicity
we have restricted to the simplified scenario described by Eq. (\ref%
{simplifiedscenario})  and we have chosen the exotic  quark Yukawa couplings
equal to unity, i.e, $y^{(T)}=y^{(J_1)}=y^{(J_2)}=1$.  In addition, the top quark
mass has been taken to be equal to $m_{t}=173$ GeV. We find that the total
cross section for the production of the $H_4$ scalar at the LHC takes a
value close to about $10^{-4}$ fb for the lower bound of $10$ TeV of the $%
SU(3)_L\times U(1)_X$ symmetry breaking scale $v_\chi$ arising from the experimental data
 on $K$, $D$ and $B$ meson mixings \cite{Huyen:2012uk} and decreases when $v_\chi$ takes larger
values. We see that the total cross  section at the LHC for the $H_4$
production via gluon fusion mechanism is small to give rise to a signal for
the allowed  values of the $SU(3)_L\times U(1)_X$ symmetry breaking scale $%
v_\chi$.  A similar situation happens at the proposed energy upgrade of the LHC with $\protect\sqrt{S}=28$ TeV, 
where this total cross section takes a value of $1.6\times 10^{-2}$fb for $v_\chi=10$ TeV as shown in 
figure \ref{ggtoH4for28TeV}. Because of the very small $H_4$ production cross section, we do not perform a detailed study of its decay modes. It is worth mentioning that the smoking gun signatures of the model under consideration will be the $Z^{\prime}$ production and the charged lepton flavor violating decay $\mu\to e\gamma$, whose observation will be crucial to assess to viability of this model.
\begin{figure}[tbh]
\resizebox{12cm}{10cm} {\includegraphics{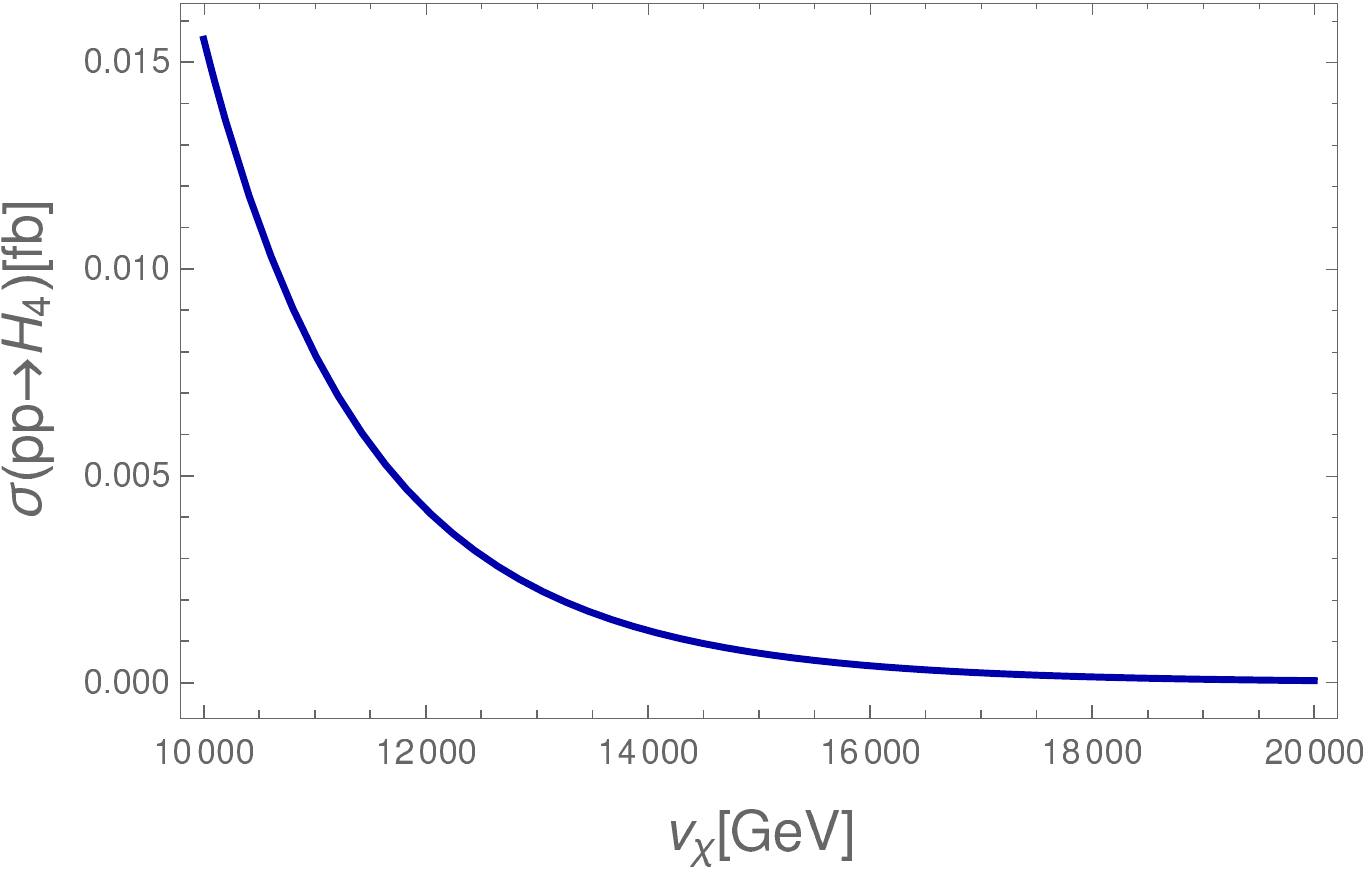}}\vspace{0cm}
\caption{Total cross section for the $H_4$ production via gluon fusion
mechanism at the proposed energy upgrade of the LHC with $\protect\sqrt{S}=28$ TeV as a function of the $SU(3)_L\times U(1)_X$ symmetry breaking scale $v_{%
\protect\chi}$ for the simplified scenario described in Eq. (\protect\ref%
{simplifiedscenario}).}
\label{ggtoH4for28TeV}
\end{figure}

\section{Dark matter relic density}
\label{DMsection}

In this section we provide a discussion of the
implications of our model for DM,  assuming that the DM
candidate is a scalar. Let us recall that our goal in this section is to
provide an estimate of the DM relic density in our model, under
some simplifying assumptions motivated by the large number of scalar fields
of the model. We do not intend to provide a sophisticated analysis of the
DM constraints of the model under consideration, which is beyond the scope of the
present paper. We just intend to show that our model can accommodate the
observed value of the DM relic density, by having a scalar DM
 candidate with a mass in the TeV range and a quartic scalar coupling
of the order unity, within the perturbative regime. We start by surveying
the possible scalar DM  candidates in the model.
Considering that the $Z_{4}$ symmetry is preserved and taking into account
the scalar assignments under this symmetry, given by Eq. (\ref%
{scalarassigments}),  we can assign this role to either any of the $%
SU( 3) _{L}$ scalar singlets, i.e.,  $Re\va^0_n$ and $Im\va^0_n$ $(n=1,2)$.
 In this work we assume that the $\va _{I}=Im \va ^{0}_1$ is the lightest among
 the  $Re\va ^0_n$ and $Im\va ^0_n$ $(n=1,2)$
  scalar fields and also lighter than the exotic charged fermions, as well as lighter than $\Psi_R$,
  thus implying that its tree-level decays
  are kinematically forbidden. Consequently, in this mass range the $Im\va ^0_1$ scalar field is stable.

The relic density is given by (c.f. Ref.~\cite{Tanabashi:2018oca,Edsjo:1997bg})
\be
\Om h^2 =\fr{0.1pb}{\left\langle \si v\right\rangle },\,\cm %
\left\langle \si v\right\rangle =\fr{A}{n_{eq}^2 }=\frac{\fr{T}{32\pi ^{4}}\dint\limits_{4m_{\va }^2 }^{\infty
}\dsum\limits_{p=W,Z,t,b,h}g_{p}^2 \fr{s\sqrt{s-4m_{\va }^2 }}{2}%
v_{rel}\si \left( \va \va \rightarrow p\overline{p}\right)
K_{1}\left( \fr{\sqrt{s}}{T}\right) ds}{\left(\frac{T}{2\pi ^{2}}\dsum\limits_{p=W,Z,t,b,h}g_{p}m_{\varphi
}^{2}K_{2}\left( \frac{m_{\varphi }}{T}\right)\right)^2}\,,
\ee%
where $\left\langle \si v\right\rangle $ is the thermally averaged
annihilation cross-section, $A$ is the total annihilation rate per unit
volume at temperature $T$ and $n_{eq}$ is the equilibrium value of the
particle density. Furthermore, $K_1$ and $K_2$ are modified Bessel functions of the second
kind and order 1 and 2, respectively \cite{Edsjo:1997bg} and $m_\va =m_{%
\func{Im}\va }$. Let us note that we assume that our scalar DM candidate is a stable weakly
interacting particle (WIMP) with annihilation cross sections mediated by
electroweak interactions mainly through the Higgs field. In addition we
assume that the decoupling of the non-relativistic WIMP of our model is
supposed to happen at a very low temperature. Because of this reason, for
the computation of the relic density, we take $T=m_{\va }/20$ as in Ref.
\cite{Edsjo:1997bg}, corresponding to a typical freeze-out temperature.
We assume that our DM candidate $\va$ annihilates mainly into $WW$,
$ZZ$, $t\overline{t}$, $b\overline{b}$ and $hh$, with annihilation cross
sections given by the following relations
\cite{Bhattacharya:2016ysw}:
\bea
v_{rel}\si \left( \va _{I}\va _{I}\rightarrow WW\right) &=&\fr{%
\la _{h^2 \va ^2 }^2 }{8\pi }\fr{s\left( 1+\fr{12m_{W}^{4}}{%
s^2 }-\fr{4m_{W}^2 }{s}\right) }{\left( s-m_{h}^2 \right)
^2 +m_{h}^2 \Ga_h^2}\sqrt{1-\fr{4m_{W}^2 }{s}},  \crn
v_{rel}\si \left( \va _{I}\va _{I}\rightarrow ZZ\right) &=&\fr{%
\la _{h^2 \va ^2 }^2 }{16\pi }\fr{s\left( 1+\fr{12m_{Z}^{4}}{%
s^2 }-\fr{4m_{Z}^2 }{s}\right) }{\left( s-m_{h}^2 \right)
^2 +m_{h}^2 \Ga _{h}^2 }\sqrt{1-\fr{4m_{Z}^2 }{s}},  \crn
v_{rel}\si \left( \va _{I}\va _{I}\rightarrow q\overline{q}%
\right) &=&\fr{N_{c}\la _{h^2 \va ^2 }^2 m_{q}^2 }{4\pi }\fr{%
\sqrt{\left( 1-\fr{4m_{f}^2 }{s}\right) ^{3}}}{\left( s-m_{h}^2 \right)
^2 +m_{h}^2 \Ga_h^2},  \crn
v_{rel}\si \left( \va _{I}\va _{I}\rightarrow hh\right) &=&\fr{%
\la _{h^2 \va ^2 }^2 }{16\pi s}\left( 1+\fr{3m_{h}^2 }{%
s-m_{h}^2 }-\fr{4\la _{h^2 \va ^2 } v^2 }{s-2m_{h}^2 }%
\right) ^2 \sqrt{1-\fr{4m_{h}^2 }{s}},
\eea%
where $\sqrt{s}$ is the centre-of-mass energy, $N_{c}=3$ is the color
factor,
$m_{h}=125.7$ GeV and $\Ga_h=4.1$ MeV are the SM Higgs boson $h$ mass
and its total decay width, respectively. Note that we have worked on the decoupling limit where
the couplings of the $126$ GeV Higgs boson to SM
particles and its self-couplings correspond to the SM expectation.

The vacuum stability and tree level unitarity constraints of the scalar potential
are \cite{EliasMiro:2012ay,Kannike:2016fmd,Cynolter:2004cq}:
\be
\la _{h^{4}}>0,\,\cm \cm \la _{\va ^{4}}>0,\,%
\cm \cm \la _{h^2 \va ^2 }^2 <\fr{2}{3}\la
_{h^{4}}\la _{\va ^{4}}.
\label{vs}
\ee
\be
\la _{\va ^4} < 8\pi ,\,\cm \cm \la
_{h^2 \va ^2 }<4\pi.
\label{us}
\ee
The dark matter relic density as a function of the
mass $m_{\va }$ of the scalar field $\va _I$ is shown in Fig.~\ref{DM}, for several values
of the quartic scalar coupling $\la _{h^2\va ^2}^2$, set to be equal to $0.7$,
$0.8$ and $0.9$ (from top to bottom). The horizontal line corresponds to the
experimental value $\Om h^2 =0.1198$ for the relic density. We found that the DM relic
density constraint gives rise to a linear correlation between the quartic scalar
coupling $\la _{h^2 \va ^2 }$ and the mass $m_{\va }$ of the
scalar DM candidate $\va _{I}$, as indicated in Fig.~\ref%
{CorrelationDM}.

We find that we can reproduce the experimental value $\Om h^2 =0.1198
\pm 0.0026$ \cite{Ade:2015xua} of the DM relic density, when the mass
$m_{\va }$ of the scalar field $\va _{I}$\
is in the range $300$ GeV$\ \lesssim m_\va \lesssim $ $570$ GeV, for a quartic scalar coupling $\la _{h^2\va^2}$ in the
window $0.5 \lesssim \la_{h^2\va^2}\lesssim 1$, which is consistent with the vacuum
 stability and unitarity constraints shown in Eqs. (\ref{vs}) and (\ref{us}).
  Note that our range of values chosen for the quartic scalar coupling $\la _{h^2\va^2}$
  also allow the extrapolation of our model at high energy scales as well as the preservation
   of perturbativity at one loop level.
\begin{figure}[t]
\center
\vspace{0cm}\includegraphics[width=0.7\textwidth]{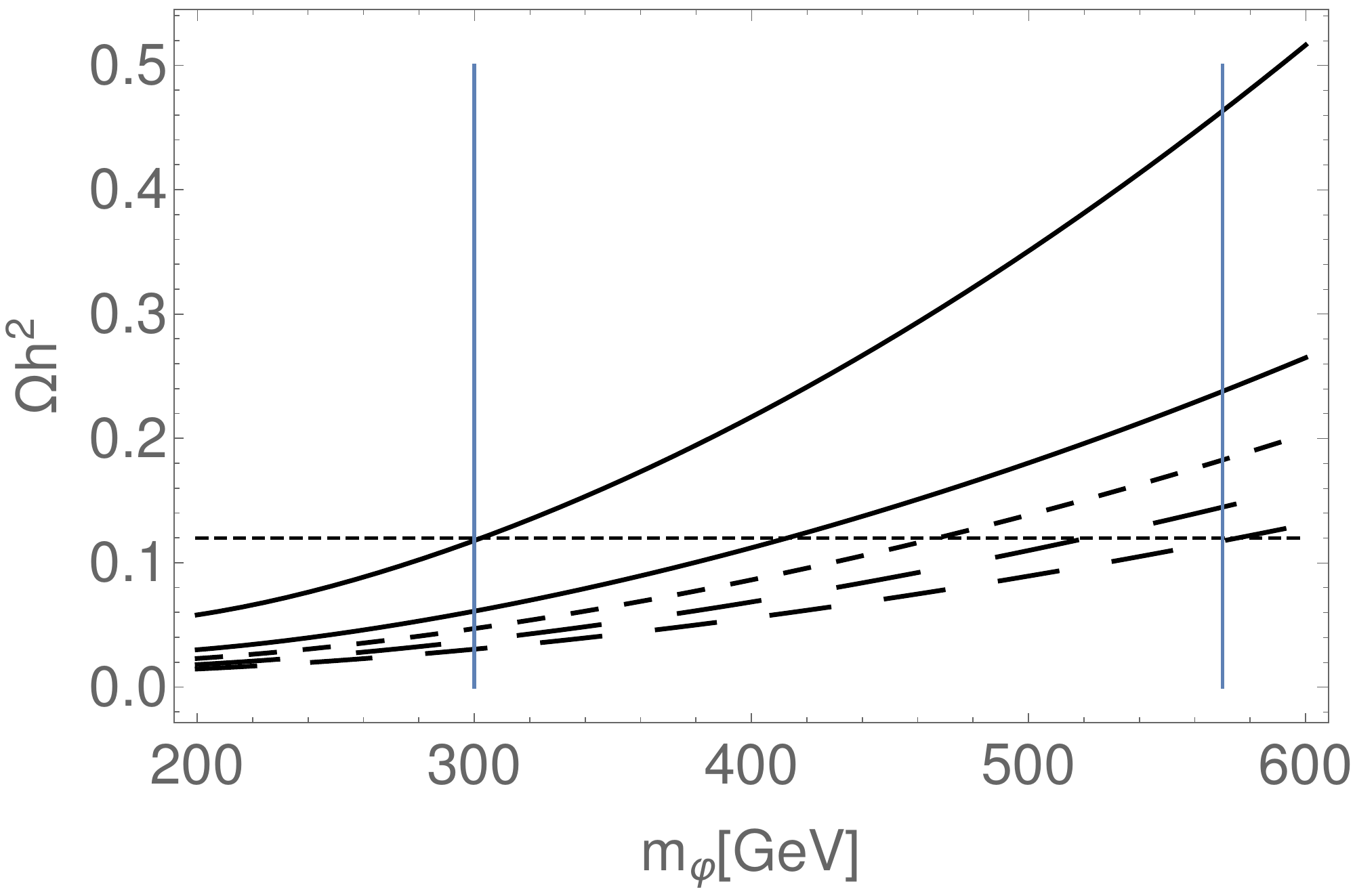}
\caption{Relic density $\Om h^2 $, as a function of the mass $m_{\protect%
\va }$ of the $\protect\va $ scalar field, for several values of the
quartic scalar coupling $\protect\la _{h^2 \protect\va ^2 }$. The
curves from top to bottom correspond to $\protect\la _{h^2 \protect%
\va^2}=0.5,0.7,0.8,0.9,1$, respectively. The horizontal line shows the
observed value $\Om h^2=0.1198$ \protect\cite{Ade:2015xua} for the
relic density. The vertical lines corresponds to the obtained lower and upper limits $300$ GeV and $570$ GeV, respectively, of the mass $m_{\protect%
\va }$ of the scalar dark matter candidate consistent with the experimental measurement of the dark matter relic density.}
\label{DM}
\end{figure}
\begin{figure}[t]
\center
\vspace{0.8cm}\includegraphics[width=0.7%
\textwidth]{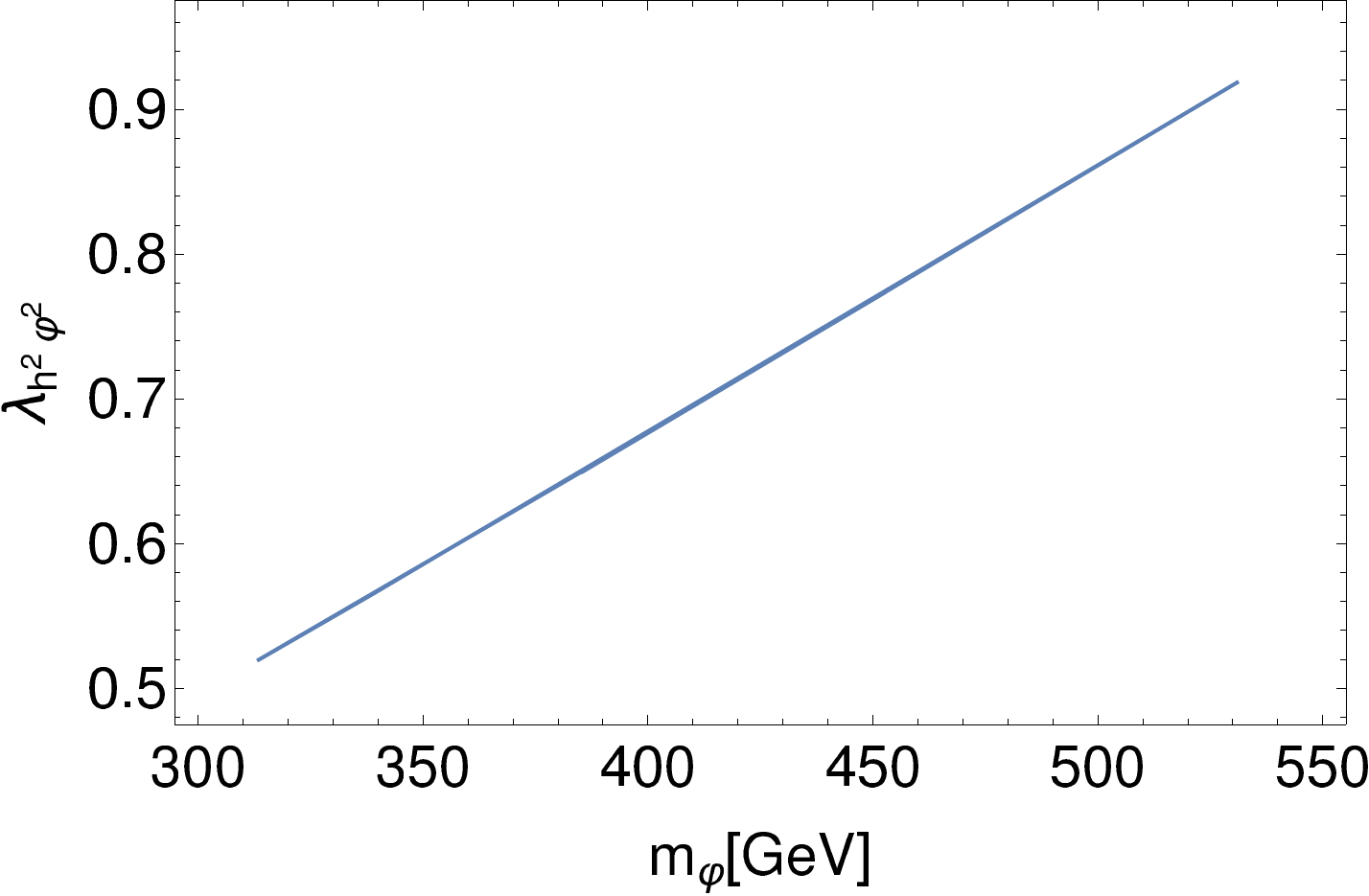}
\caption{Correlation between the quartic scalar coupling and the mass $m_{%
\protect\va }$ of the scalar DM  candidate $\protect\va $,
consistent with the experimental value $\Om h^2=0.1198$ for the Relic
density.}
\label{CorrelationDM}
\end{figure}

\section{Conclusions}
\label{conclusions}
We have studied some phenomenological aspects of the extended inert 331 model, which incorporates the mechanism of sequential loop-generation of the SM fermion masses, explaining the observed strong hierarchies between them as well as the corresponding mixing parameters. 
A particular emphasis has been made on analyzing the constraints arising from the experimental data on the $\rho$ parameter, as well as those ones resulting from the charged lepton flavor violating process $\mu\to e\gamma$ and dark matter. Furthermore, we have studied the production of the heavy $Z^{\prime }$ gauge boson 
in proton-proton collisions via the Drell-Yan mechanism. We found that the corresponding total cross section at the LHC ranges from $85$ fb up to $10$ fb when the $Z^\prime $ gauge boson mass is varied within $4-5$ TeV interval. The $Z^{\prime }$ production cross section gets significantly enhanced at the proposed energy upgrade of the LHC with $\sqrt{S}=28$ TeV reaching the typical values of $2.5-0.7$ pb. From these results we find that the $pp\to Z^\prime\to l^{+}l^{-}$ resonant production cross section reach values of about $1$ fb and $10^{-2}$ pb at $M_{Z^{\prime }}=4$ TeV, for $\protect\sqrt{S}=14$ TeV and $\protect\sqrt{S}=28$ TeV, respectively. 
These obtained values for the $pp\to Z^\prime\to l^{+}l^{-}$ resonant production cross sections are below and of the same order of magnitude of its corresponding lower experimental limit arising from LHC searches, for $\sqrt{S}=13$ TeV and $\sqrt{S}=28$ TeV, respectively.  
 Besides that, we have found that $Z'$ gauge bosons heavier than about $4$ TeV comply with the experimental constraints on the oblique $\rho$ parameter as well as with the collider constraints. 
In addition, we have found that the constraint on the charged lepton flavor violating decay $\mu\rightarrow e\gamma$ set the sterile neutrino masses to be lighter than about $1.12$ TeV. We have found that the obtained values of the branching ratio for the $\mu\to e\gamma$ decay are located in the range $3\times 10^{-13}\lesssim
Br\left( \mu \rightarrow e\gamma \right) \lesssim 4\times 10^{-13}$, whereas the obtained branching ratios for the $\tau
\rightarrow \mu \gamma $ and $\tau \rightarrow e\gamma $ decays can reach
values of the order of $10^{-13}$. Consequently, our model predicts charged lepton flavor violating decays within the reach of future experimental
sensitivity. 
We found that the total cross section for the production of the $H_4$ scalar
at the LHC with $\sqrt{S}=13$ TeV takes a value close to about $10^{-4}$ fb for
the lower bound of $10$ TeV of the $SU(3)_L\times U(1)_X$ symmetry
breaking scale $v_\chi$ required by the consistency  of the $\rho$ parameter with
the experimental data. This value is increased to $1.6\times 10^{-2}$ fb at the proposed energy upgrade of the LHC with $\protect\sqrt{S}=28$ TeV, thus implying that the total cross section at the LHC for the $H_4$ production via gluon fusion mechanism is very small to give rise to a signal for
the allowed  values of the $SU(3)_L\times U(1)_X$ symmetry breaking scale $%
v_\chi$ even at the proposed energy upgrade of the LHC. We also analyzed in detail the scalar potential and the gauge sector of the model.  

The general Higgs sector is separated into two parts. The first part
consists of lepton number conserving terms and the second one contains
lepton number violating couplings. The first part of potential was considered in
details and the SM Higgs boson was derived and as expected, mainly arises from $\eta_1^0$. We have showed that the whole scalar potential,
excepting its CP-even sector, has a quite similar situation since the resulting physical scalar mass spectrum is similar in both cases. The scalar spectrum contains enough number of Goldstone
bosons for massive gauge bosons. In the CP-odd scalar sector, there are four massive bosons and one of them
is a DM candidate. The CP-even scalar mass spectrum consists of seven massive fields including the SM Higgs boson and a DM candidate.
The singly electrically charged Higgs boson sector contains six massive fields. Two of them have masses at the electroweak scale and
the remaining one has a mass around $3.5$ TeV. The masses for the three charged bosons $\phi_i^+, i = 1,2,4$ are not fixed.
The scalar potential contains a Majoron but it is
harmless, because it is a scalar singlet. Due to the unbroken $Z_{4}$ symmetry our model has the stable scalar dark matter candidates $Re\va^0_n$ and $Im\va^0_n$ $(n=1,2)$ and the fermionic dark matter candidate $\Psi_R$. In this work we assume that $\va _{I}=Im \va ^{0}_1$ is the lightest among
 the $Re\va ^0_n$ and $Im\va ^0_n$ $(n=1,2)$
  scalar fields and also lighter than the exotic charged fermions and than $\Psi_R$, which implies that it is stable and thus it is the dark matter candidate considered in this work.  
To reproduce the Dark matter relic density, the mass of the scalar dark matter candidate has
 to be in the range $300$ GeV$\ \lesssim m_{\va }\lesssim 570$ GeV,  for
 a quartic scalar coupling $\la _{h^2 \va ^2 }$ in the window $0.5\lesssim \la _{h^2\va ^2}\lesssim 1$.
 In addition, it has been shown in Ref. \cite{CarcamoHernandez:2017cwi} that requiring that the DM candidate $\va^{0}$
 lifetime be greater than the universe lifetime
$\tau_{u} \approx 13.8$ Gyr and assuming $m_{\va^{0}} \sim 1$ TeV, we estimate the cutoff scale of our
model $\La > 3\times 10^{10}$ GeV. Thus we conclude that under the above specified conditions the model
 contains viable fermionic $\Psi_{R}$ and scalar $\va^{0}$ DM candidates. A sophisticated analysis
 of the DM constraints of our model is beyond the scope of the present paper and is left for future studies.

\section*{Acknowledgments}
This research has been financially supported by Fondecyt (Chile), Grants
No.~1170803, CONICYT PIA/Basal FB0821, the Vietnam National Foundation for
Science and Technology Development (NAFOSTED) under grant number
103.01-2017.356. 
H. N. L.  is very grateful  to the Bogoliubov Laboratory for Theoretical Physics, JINR, Dubna, Russia  for the warm hospitality during his visit.

\appendix
\section{The scalar potential}
\label{scalarpotential}
The renormalizable potential contain three parts: the
first one
invariant under group $\mathcal{G}$ in (\ref{eq0}) is given by
\bea
V_{LNC}&=& \mu^2_\chi \chi^\dag \chi+ \mu^2_\rho \rho^\dag \rho + \mu^2_\eta
\eta^\dag \eta + \sum_{i=1}^{4}  \mu^2_{\phi_i^+}  \phi_i^+
\phi_i^- + \sum_{i=1}^2  \mu^2_{\va_i} \va_i^0 \va_i^{0*} +
\mu^2_\xi \xi^{0*}\xi^0  \crn
&+&\chi^\dag \chi ( \la_{13} \chi^\dag \chi + \la_{18}\rho^\dag \rho
+ \la_5 \eta^\dag \eta ) + \rho^\dag \rho(\la_{14} \rho^\dag \rho
+\la_6 \eta^\dag \eta) + \la_{17} (\eta^\dag \eta)^2  \crn
&+& \la_{7} (\chi^\dag \rho)(\rho^\dag \chi) + \la_{8} (\chi^\dag
\eta)(\eta^\dag \chi) + \la_{9}(\rho^\dag \eta) (\eta^\dag \rho)  \crn
&+&\chi^\dag \chi \left(\sum_{i=1}^{4} \la^{\chi \phi}_{i} \phi_i^+
\phi_i^- + \sum_{i=1}^2  \la^{\chi \va}_{i} \va_i^0
\va_i^{0*} + \la_{\chi \xi} \xi^{0*}\xi^0 \right)  \crn
&+&\rho^\dag \rho \left(\sum_{i=1}^{4} \la^{\rho \phi}_{i} \phi_i^+
\phi_i^- + \sum_{i=1}^2  \la^{\rho \va}_{i} \va_i^0
\va_i^{0*} + \la_{\rho \xi} \xi^{0*}\xi^0 \right)  \crn
&+&\eta^\dag \eta \left(\sum_{i=1}^{4} \la^{\eta \phi}_{i} \phi_i^+
\phi_i^- + \sum_{i=1}^2  \la^{\eta \va}_{i} \va_i^0
\va_i^{0*} +\la_{\eta \xi} \xi^{0*}\xi^0 \right)  \crn
&+& \sum_{i=1}^{4} \phi_i^+ \phi_i^-\left( \sum_{j=1}^{4} \la^{\phi
\phi}_{ij} \phi_j^+ \phi_j^- + \sum_{j=1}^2  \la^{\phi \va}_{ij}
\va_j^0 \va_j^{0*} +\la_i^{\phi \xi} \xi^{0*}\xi^0 \right)
\crn
&+& \sum_{i=1}^2  \va_i^0 \va_i^{0*} \left(
\sum_{j=1}^2 \la^{\va \va}_{ij} \va_j^0 \va_j^{0*} +
\la^{\va\xi}_{i} \xi^{0*}\xi^0\right) +\la_\xi ( \xi^{0*}\xi^0)^2
\crn
&+& \left\{ \la_{10}\left(\phi_2^+\right)^2 \left(\phi_3^-\right)^2
+\la_{11}\left(\phi_2^+\right)^2 \left(\phi_4^-\right)^2
+\la_{12}\left(\phi_3^+\right)^2 \left(\phi_4^-\right)^2\right.  \notag
\\
&+ & w_{1}\left( \va_2 ^{0}\right)^2 \va _{1}^{0}+ w_2  \chi
^\+ \rho\phi _{3}^{-} + w_{3}\eta^{\dagger}\chi\xi^0+ w_{4}\left(
\va_2 ^{0}\right)^2 \va _{1}^{0\ast}+w_{5}\phi_{3}^{+}\phi_{4}^{-}
\va _{1}^{0} +w_{6}\phi_{3}^{+}\phi_{4}^{-} \va _{1}^{0\ast}  \notag
\\
&+& \chi\rho\eta (\la_{1} \va _{1}^{0} +\la_2  \va
_{1}^{0\ast})+\chi^\dagger\rho\phi_4^-\left(\la_{15} \va _{1}^{0} +
\la_{16}\va _{1}^{0*}\right)+ \la_{3}\eta^{\dagger}\rho\phi
_{3}^{-} \xi^0+\la_{4}\phi _{1}^{+}\phi _2 ^{-} \va _2 ^{0} \xi^{0}
\crn
&+& \left( \la_{19}\phi_3^- \phi_4^+ +\la_{20}\phi_3^+ \phi_4^-
\right) \left( \va _2 ^{0}\right) ^2  +\la_{21}\left(
\va_{1}^{0}\right)^{3}\va_{1}^{0\ast}  \crn
&+& \left(\la_{22}\chi^\dag \chi +\la_{23}\rho^\dag \rho
+\la_{24}\eta^\dag \eta + \sum_{i=1}^{4} \la_{61i} \phi_i^+ \phi_i^-
+ \sum_{i=1}^2  \la_{62i} \va_i^0 \va_i^{0*}\right.  \crn
&+&\left. \left. \la_{25}\xi^{0*}\xi^0 \right)(\va_1^{0})^2 +h.c.
\right\}  \label{ct2}
\eea

The second part is a lepton number violating one (the subgroup $U(1)_{L_g}$
is violated)
\bea
V_{LNV}&=& \mu_{\chi\eta}^2 \left(\chi ^\+ \eta+\eta ^{\dagger
}\chi\right) +\left[\la _{26}(\chi ^\+ \chi)+\la _{27}(\rho
^\+ \rho)+\la _{28}(\eta^\+ \eta) \right](\chi ^{\dagger
}\eta+\eta ^\+ \chi)  \crn
& + &\la _{29}\left[(\chi ^\+ \eta)^2 +(\eta ^\+ \chi)^2%
\right] + \la_{30}\left[(\eta^\+ \rho )(\rho ^\+ \chi )
+(\chi^\+ \rho )(\rho ^\+ \eta) \right]  \crn
&+& \left\{ \xi^0 \left( w_{7}\chi^\dag \chi +w_{8}\rho^\dag \rho
+w_{9}\eta^\dag \eta + \sum_{i=1}^{4} w_{2i} \phi_i^+ \phi_i^- +
\sum_{i=1}^2  w_{3i} \va_i^0 \va_i^{0*} +w_{10}\xi^{0*}\xi^0
\right) \right.  \crn
&+& \xi^0 \left[ w_{11}\chi^\dag \eta +w_{12}\left(\va_1^0\right)^2
+w_{13}\left(\va_1^{0*}\right)^2 +w_{14}\left(\xi^0\right)^2 \right]
+w_{15}\eta^\dag\rho\phi_3^- +w_{16}\phi_2^- \phi_1^+ \va_2^0  \crn
&+& \!\!\left(\xi^0\right)^2  \left[ \la_{31}\chi^\dag \chi
+ \la_{32}\rho^\dag \rho + \la_{33}\eta^\dag \eta +
\sum_{i=1}^{4} \la_{63i} \phi_i^+ \phi_i^- +\!\! \sum_{i=1}^2
\la_{64i} \va_i^0 \va_i^{0*}
+ \la_{34}\xi^{0*}\xi^0\right.  \crn
& +& \left. \la_{35}\left(\va_1^0\right)^2
+ \la_{36}\left(\va_1^{0*}\right)^2 \right]  \crn
&+& \!\chi^\dag \eta \left[ \sum_{i=1}^{4} \la_{65i} \phi_i^+ \phi_i^-
+  \sum_{i=1}^2  \la_{66i} \va_i^0 \va_i^{0*}
+ \la_{37}\xi^{0*}\xi^0 +\!\la_{38}\left(\va_1^0\right)^2
+ \la_{39}\left(\va_1^{0*}\right)^2\right.  \crn
& +& \left. \la_{40}\left(\xi^0\right)^2
+ \la_{41}\left(\xi^{0*}\right)^2 \right]  \crn
&+& \eta^\dag \rho \left(\la_{42}\phi_4^- \va_1^0
+\la_{43}\phi_4^- \va_1^{0*} +\la_{44}\phi_3^- \xi^{0*}\right)
+\rho^\dag \chi \phi_3^+ \left(\la_{45}\xi^0+\la_{46}\xi^{0*}\right)
\crn
&+& \la_{47}\left(\phi_1^+\right)^2 \phi_3^- \phi_4^- + \phi_1^+
\phi_2^- \left(\la_{48} \va_1^0 \va_2^{0*}
+\la_{49}\va_1^{0*} \va_2^{0*}+\la_{50}\va_2^0
\xi^{0*}\right)  \crn
&+& \phi_3^+ \phi_4^- \left( \la_{51}\va_1^0 \xi^0
+\la_{52}\va_1^0 \xi^{0*} +\la_{53}\va_1^{0*} \xi^0
+\la_{54}\va_1^{0*} \xi^{0*}\right)  \crn
&+& \left. \left(\va_2^0\right)^2 \left( \la_{55}\va_1^0 \xi^0
+\la_{56}\va_1^0 \xi^{0*} +\la_{57}\va_1^{0*} \xi^0
+\la_{58}\va_1^{0*} \xi^{0*}\right) \;+\; h.c. \right\}  \label{VLNV}
\eea

The last part which breaks softly $Z_{4}\times Z_2 $, is given by
\bea
L_{g soft}^{scalars} &=& \mu_4^2 \va_1^0 \va_2^0 +\mu_6^2
\va_1^0 \va_2^{0*} + \mu_1^2 \left(\va_2^0\right)^2 +\mu_2^2
\phi_2^+\phi_3^- +\mu_5^2 \phi_2^+ \phi_4^- + \mu_3^2 \phi_3^+ \phi_4^- +h.c.
\label{eq7161}
\eea

The total potential is composed of three above mentioned parts
\be
V=V_{LNC}+V_{LNV} +\mathcal{L}_{soft}^{scalars}\, .  \label{eq7162}
\ee
The scalar interactions needed for quark and charged lepton mass generation,
read as follows
\be
L_{Higgsqcl} = \la_{1} \chi\rho\eta \va _{1}^{0} +
\la_{3}\eta^{\dagger}\rho\phi _{3}^{-} \xi^0 +\la_{4}\phi
_{1}^{+}\phi _2 ^{-} \va _2 ^{0} \xi^{0} +w_{1}\left(
\va_2 ^{0}\right)^2 \va _{1}^{0}+ w_2  \chi ^\+ \rho\phi
_{3}^{-}+h.c \, .  \label{eq7163}
\ee
For the neutrino mass generation, beside the first term in (\ref{eq7163}),
the additional part is given as
\be
L_{Higgsneutrino} = \la_{13}(\chi^\dag \chi)^2 + \la_{5}(\chi^\dag
\chi)(\eta^\dag \eta) +\left[\la _{27}(\rho ^\+ \rho)(\chi
^\+ \eta+\eta ^\+ \chi) +\mu _{3}^2 \phi _{4}^{-}\phi
_{3}^{+}+h.c \right]\, .  \label{eq7164}
\ee
It is worth mentioning that for the generation of masses for quark and charged
lepton, only terms in the conserving part $V_{LNC}$ are enough, while for the
generation of the light active neutrino masses, one needs the lepton number violating scalar interactions of $V_{LNV}$ as well as
 the softly breaking part $\mathcal{L}%
_{soft}^{scalars}$ [the last term in (\ref{eq7164})] of the scalar potential.

\end{document}